\documentclass[journal]{IEEEtran}

\usepackage{amsmath}

\usepackage{cite}

\usepackage{latexsym}
\usepackage{graphics}
\usepackage{amssymb}
\usepackage{amsthm}

\usepackage{cite}
\usepackage{array}
\usepackage{mdwmath}
\usepackage{mdwtab}
\usepackage[tight,footnotesize]{subfigure}
\usepackage{graphicx}

\usepackage{stfloats}
\fnbelowfloat
\usepackage{url}

\ifCLASSOPTIONcompsoc
\usepackage[tight,normalsize,sf,SF]{subfigure}
\else
\usepackage[tight,footnotesize]{subfigure}
\fi

\newtheorem{example}{Example}

\def\n0{n_{0}}

\def\tr{\mathrm{tr}}

\def\rank{\mathrm{rank}}

\newcommand{\fracSum}[1]{{\underset{{#1}}{\sum}}}

\newcommand{\vect}[1]{\mathbf{#1}}

\newcommand{\maximize}[1]{{\underset{{#1}}{\mathrm{maximize}}}}
\newcommand{\minimize}[1]{{\underset{{#1}}{\mathrm{minimize}}}}

\theoremstyle{remark}

\newtheorem{theorem}{Theorem}
\newtheorem{corollary}{Corollary}
\newtheorem{lemma}{Lemma} 
\newtheorem{definition}{Definition}

\newcommand{\mysmallarraydecl}{\renewcommand{%
\IEEEeqnarraymathstyle}{\scriptscriptstyle}%
\renewcommand{\IEEEeqnarraytextstyle}{\scriptsize}%
\renewcommand{\baselinestretch}{1.1}%
\settowidth{\normalbaselineskip}{\scriptsize
\hspace{\baselinestretch\baselineskip}}%
\setlength{\baselineskip}{\normalbaselineskip}%
\setlength{\jot}{0.25\normalbaselineskip}%
\setlength{\arraycolsep}{2pt}}

\newenvironment{lined}[1]%
 {\begin{center}\begin{minipage}{#1}\hrule\medskip}
 {\vspace{-1ex}\hrule \end{minipage}\end{center}}

\begin{document}

\title{Optimality Properties, Distributed Strategies, and
Measurement-Based Evaluation of Coordinated Multicell OFDMA
Transmission}

\author{Emil Bj\"ornson,~\IEEEmembership{Student Member,~IEEE,}
        Niklas Jald\'en,~\IEEEmembership{Member,~IEEE,}
        Mats Bengtsson,~\IEEEmembership{Senior Member,~IEEE,}
        and\\ Bj\"orn~Ottersten,~\IEEEmembership{Fellow,~IEEE}
\thanks{Manuscript received July 13, 2010; revised February 21, 2011 and June 17,
2011; accepted August 01, 2011. Date of publication August 22, 2011; date
of current version November 16, 2011. The associate editor coordinating the
review of this manuscript and approving it for publication was Prof. Gerald
Matz. This work was supported by the European Research Council under the
European Communitys Seventh Framework Programme (FP7/2007-2013)/ERC
grant agreement number 228044.}%
\thanks{E. Bj\"ornson and M. Bengtsson are with the Signal Processing Laboratory,
ACCESS Linnaeus Center, KTH Royal Institute of Technology, SE-100 44
Stockholm, Sweden (e-mail: emil.bjornson@ee.kth.se; mats.bengtsson@ee.kth.se).}%
\thanks{N.~Jald\'en is with Ericsson Research, SE-164 80 Stockholm, Sweden (e-mail:
niklas.jalden@ericsson.com).}%
\thanks{B. Ottersten is with Signal Processing Laboratory, ACCESS Linnaeus
Center, KTH Royal Institute of Technology, SE-100 44 Stockholm, Sweden.
He is also with Interdisciplinary Centre for Security, Reliability
and Trust, University of Luxembourg, L-1359 Luxembourg-
Kirchberg, Luxembourg (e-mail: bjorn.ottersten@ee.kth.se).}%
\thanks{Digital Object Identifier 10.1109/TSP.2011.2165706}%
}

\markboth{IEEE TRANSACTIONS ON SIGNAL PROCESSING, VOL.~59, NO.~12, DECEMBER 2011}%
{Bj\"ornson \MakeLowercase{\textit{et al.}}: IEEE TRANSACTIONS ON
SIGNAL PROCESSING}

\maketitle

\begin{abstract}
The throughput of multicell systems is inherently limited by
interference and the available communication resources. Coordinated
resource allocation is the key to efficient performance, but the
demand on backhaul signaling and computational resources grows
rapidly with number of cells, terminals, and subcarriers. To handle this, we propose a
novel multicell framework with dynamic cooperation clusters where each terminal is jointly served by a small set of base stations.
Each base station coordinates interference to neighboring terminals only, thus limiting backhaul signalling and making the framework scalable.
This framework can describe anything from interference channels to ideal joint multicell transmission.

The resource allocation (i.e., precoding and scheduling) is formulated as an optimization problem \eqref{eq_P1}
with performance described by arbitrary monotonic functions of the signal-to-interference-and-noise ratios
(SINRs) and arbitrary linear power constraints. Although \eqref{eq_P1} is non-convex and difficult to solve optimally, we are able to prove:
1) Optimality of single-stream beamforming; 2) Conditions for full power usage; and 3) A precoding parametrization based on a few parameters between zero and one.
These optimality properties are used to propose low-complexity strategies: both a centralized scheme and a distributed version that only requires local channel knowledge and processing. We evaluate the performance on measured multicell channels and observe that the proposed strategies achieve close-to-optimal performance among centralized and distributed solutions, respectively.
In addition, we show that multicell interference coordination can give substantial improvements in sum performance, but that joint transmission is very sensitive to synchronization errors and that some terminals can experience performance degradations.
\end{abstract}


\IEEEpeerreviewmaketitle

\section{Introduction}
\IEEEPARstart{I}{n} conventional cellular systems, each terminal belongs to one cell
at a time and data transmission is scheduled autonomously by its
base station. We consider systems where each base station can divide
its transmit resources between terminals using orthogonal
frequency-division multiple access (OFDMA), which generates
independent subcarriers \cite{Jiang2007a}. In addition, multiple
terminals can be assigned to each subcarrier using space division multiple access (SDMA)
and multiple-input multiple-output (MIMO) techniques that
manage co-terminal interference within the cell \cite{Gesbert2007a}.
However, with base stations performing autonomous \emph{single-cell
processing}, the performance is fundamentally limited by
interference from other cells---especially for terminals close to
cell edges.

The limiting inter-cell interference can be handled by base station
coordination, recently termed \emph{network MIMO}
\cite{Venkatesan2007a} and \emph{coordinated multi-point
transmission} (CoMP) \cite{Parkvall2008a}. By sharing data and
channel state information (CSI) over the backhaul, base stations can
coordinate the interference caused to adjacent cells, and cell edge
terminals can be jointly served through multiple base stations
\cite{Shamai2001a,Zhang2004a,Karakayali2006a}. The multicell
capacity was derived in \cite{Weingarten2006a}, while more practical
performance gains over conventional single-cell processing were
reported in \cite{Marsch2008a, Simeone2009a, Bjornson2009e} under
constrained backhaul signaling. In practice, the transmission
optimization is also constrained by computational complexity and the difficulty of obtaining reliable CSI, which
makes centralized implementations of multicell coordination
intractable in large networks \cite{Bjornson2009e}.

In practical multicell systems, only a small subset of base
stations will take the interference generated at a given terminal
into consideration (to limit the backhaul signaling and complexity,
and to avoid estimating negligibly weak channels). Fixed cooperation
clusters were considered in \cite{Marsch2008a} to (iteratively)
coordinate transmissions within each cluster. While easily
implementable for co-located base stations (such as sectors
connected to the same eNodeB in an LTE system), the performance is
still limited by out-of-cluster interference.
A more dynamic approach was taken in \cite{Tolli2008a} where base stations serve partially
overlapping sets of terminals. Still, global interference coordination was assumed, making the approach infeasible in large systems.

Resource allocation is very difficult to solve optimally, even under simplifying assumptions such as a single subcarrier, global interference coordination, and perfect CSI. A few type of problems can be solved using \emph{uplink-downlink duality} \cite{Boche2002a,Wiesel2006a,Yu2007a,Dahrouj2010a},
but weighted sum rate optimization and similar problems are all NP-hard \cite{Liu2011a}. There are algorithms
that find the optimal solutions \cite{Brehmer2010a,Bjornson2012a}, but with unpractically slow convergence. Suboptimal iterative solutions
that perform well on synthetic channels have been suggested in \cite{Tolli2009b,Zheng2008a,Tolli2008a,Venturino2010a}.
However, synthetic and real channels usually differ due to the simplifications used in the channel model assumptions.
The performance of downlink multicell coordination has
not been evaluated on measured channels, thus characteristics such
as the correlation between channels from different base stations
have not been considered \cite{Jalden2007a}.

Herein, we analyze coordinated multicell OFDMA transmission, derive properties of the
optimal resource allocation, propose low-complexity strategies, and analyze the
performance on measured channels. The major contributions are:

\begin{itemize}
\item We propose a general multicell cooperation framework that enables unified analysis of anything from interference channels to ideal network MIMO. The main characteristic is that each base station is responsible for the interference caused to a set of terminals, while only serving a subset of them with data (to limit backhaul signaling).

\item Multicell OFDMA transmission and resource allocation is formulated as an optimization
problem \eqref{eq_P1} with arbitrary monotonic utility functions
(e.g., representing data rates, error rates, or mean square
errors), single user detection, and arbitrary linear power constraints.

\item Three properties of the optimal solution to \eqref{eq_P1} are derived: 1) Optimality of single-stream beamforming; 2) Conditions for full power usage; 3) An explicit precoding parametrization based on a few real-valued parameters between zero and one. This novel parametrization improves prior work in \cite{Jorswieck2008b,Bjornson2010c,Shang2010a,Zhang2010a,Mochaourab2011a} by supporting general multicell and multicarrier systems and generally requiring much fewer parameters.

\item Two low-complexity strategies for resource allocation are proposed based on the three optimality properties and an efficient algorithm for subcarrier allocation.
The centralized strategy provides close-to-optimal performance, while the distributed version is suitable for large
systems with many subcarriers and where the backhaul and computational
resources required for the iterative solutions in
\cite{Tolli2009b,Zheng2008a,Tolli2008a,Venturino2010a} are unavailable.

\item As the performance of any communication system depends on the
channel characteristics, realistic conditions are necessary for
reliable evaluation. Therefore, the proposed strategies are
evaluated on measured channel vectors from a typical urban
macro-cell environment. The impact of multicell coordination is
evaluated both in terms of average performance and for fixed terminal
locations, and the robustness to synchronization imperfections is studied.
\end{itemize}

Preliminary single-carrier results were reported in
\cite{Bjornson2010d}.

\vskip 3mm
\textbf{Notation:} $\vect{X}^T$, $\vect{X}^H$, and
$\vect{X}^{\dagger}$ denote the transpose, the conjugate transpose,
and the Moore-Penrose inverse of $\vect{X}$, respectively.
$\vect{I}_N$ and $\vect{0}_N$ are $N \times N$ identity and zero
matrices, respectively. If $\mathcal{S}$ is a set, then its members
are $\mathcal{S}(1),\ldots,\mathcal{S}(|\mathcal{S}|)$ where
$|\mathcal{S}|$ is the cardinality.

\section{General Multicell System Model}

We consider a downlink system with $K_t$ transmitters, where transmitter $j$ is equipped with $N_j$ antennas. They communicate over $K_c$
independent subcarriers with $K_r$ receivers having one effective
antenna each.\footnote{This model applies to multi-antenna receivers
that fix their linear receivers prior to transmission optimization. This case is relevant both for low-complexity transceiver design and as part of iterative transmitter/receiver optimization algorithms.} The
transmitters and receivers are denoted $\textrm{BS}_j$ and
$\textrm{MS}_k$, respectively, for $j \in \mathcal{J} =
\{1,\ldots,K_t \}$ and $k \in \mathcal{K} = \{1,\ldots,K_r \}$.

In a general multicell scenario, some terminals are served in a coordinated manner by multiple transmitters. In addition, some transmitters and
receivers are very far apart, making it impractical to estimate and
separate the interference on these channels from the noise. Based on these observations, we propose a general multicell coordination framework:

\begin{definition} \label{def_dynamic_clusters}
\emph{Dynamic cooperation clusters} means that $\textrm{BS}_j$
\begin{itemize}
\item Has channel estimates to receivers in $\mathcal{C}_j
\subseteq \{1,\!...,\!K_r\!\}$, while interference generated to receivers $\bar{k} \not \in \mathcal{C}_j$ is negligible and can be treated as background noise;\footnote{This means that $\textrm{BS}_j$ has CSI to all users that receive non-negligible interference from $\textrm{BS}_j$---a natural assumption since these are the users where $\textrm{BS}_j$ can achieve reliable channel estimates. But compared with autonomous single-cell processing, it requires additional estimation, feedback, and backhaul resources not necessarily available in all system architectures.}
\item Serves the receivers in $\mathcal{D}_j \subseteq
\mathcal{C}_j$ with data.
\end{itemize}
\end{definition}

\begin{figure*}[t!]
\begin{center}
\includegraphics[width=1.3\columnwidth]{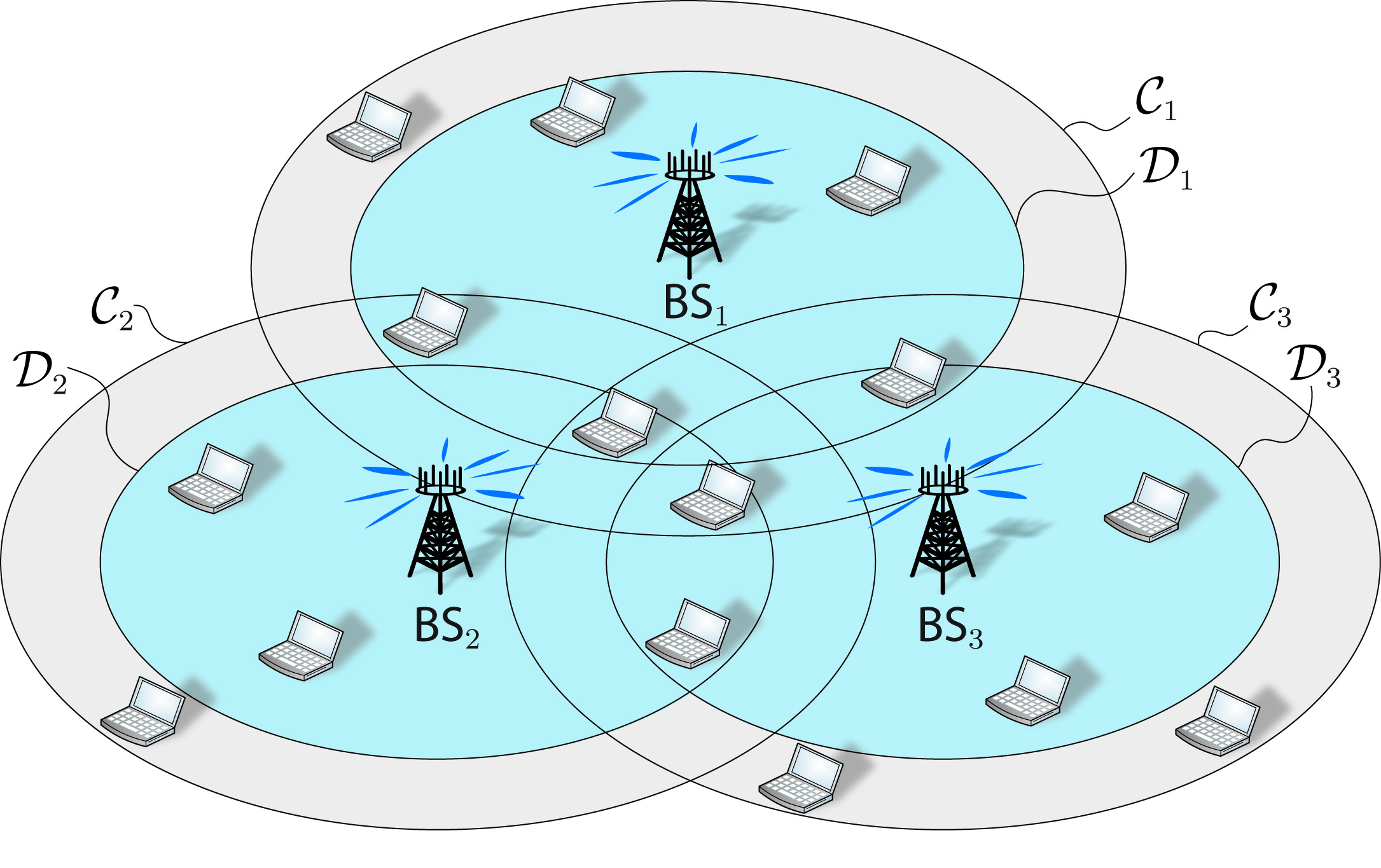}
\end{center} \vskip-3mm
\caption{Intersection between three cells. $\textrm{BS}_j$ serves terminals in the inner circle ($\mathcal{D}_j$) and coordinates interference to terminals in the outer circle ($\mathcal{C}_j$). Ideally, negligible interference is caused to terminals outside both circles.}\label{figure_system_model}
\end{figure*}

This coordination framework is characterized by the sets $\mathcal{C}_j,\mathcal{D}_j$ for all $j \in \mathcal{J}$, which are illustrated in Fig.~\ref{figure_system_model}.
The mnemonic rule is that $\mathcal{D}_j$ describes \emph{data} from transmitter $j$, while $\mathcal{C}_j$ describes \emph{coordination} from transmitter $j$.
How to select these sets efficiently, by only accepting the overhead involved in interference coordination and joint transmission if the expected performance gains are substantial, is a very interesting and important problem. The solution depends on the system architecture and is beyond the scope of
this paper (see \cite{Zhang2004b,Fuchs2006a,Parkvall2008a,Xu2010a}), but a simple scheme would be to include $\textrm{MS}_k$ in $\mathcal{C}_j$ if the long-term average channel gain from $\textrm{BS}_j$ is above a certain threshold. If the gain is above a second threshold, the terminal is also included in $\mathcal{D}_j$. In principle, different $\mathcal{C}_j,\mathcal{D}_j$ can be used for each subcarrier, but it is not necessary since subcarrier allocation naturally appears in the resource allocation analyzed herein. Observe that although $\mathcal{C}_j,\mathcal{D}_j$ can be selected decentralized at transmitter $j$, some mechanism for coordination and data sharing is required between adjacent transmitters.

At subcarrier $c$, the channel from $\textrm{BS}_j$ to
$\textrm{MS}_k$ is denoted $\vect{h}_{jkc} \in \mathbb{C}^{N_j
\times 1}$. The collective channel from all transmitters is $\vect{h}_{kc}=[\vect{h}_{1kc}^T
\ldots \vect{h}_{K_t kc}^T ]^T \in \mathbb{C}^{N}$, where $N=\sum_{j \in \mathcal{J}} N_j$ is the total number of transmit antennas. Based on the framework in
Definition \ref{def_dynamic_clusters}, only certain channel elements in $\vect{h}_{kc}$ will carry data and/or non-negligible interference. These can be selected
by the diagonal matrices $\vect{D}_k \in
\mathbb{C}^{N \times N}$ and $\vect{C}_k  \in
\mathbb{C}^{N \times N}$ defined as
\begin{equation}
\vect{D}_{k}= \left[\begin{IEEEeqnarraybox*}[][c]{c,c,c}
\vect{D}_{1k} &  & \vect{0}\\[-1.2ex]
 & \ddots &  \\[-1.2ex]
\vect{0} &  & \vect{D}_{K_t k}
\end{IEEEeqnarraybox*}  \right] \quad \textrm{where} \quad \vect{D}_{jk}= \begin{cases}
\vect{I}_{N_j}, & \textrm{if } k \in \mathcal{D}_j,\\[-1.2ex]
\vect{0}_{N_j}, & \textrm{otherwise},
\end{cases}
\end{equation}
\begin{equation}
\vect{C}_{k}= \left[\begin{IEEEeqnarraybox*}[][c]{c,c,c}
\vect{C}_{1k} &  & \vect{0}\\[-1.2ex]
 & \ddots &  \\[-1.2ex]
\vect{0} &  & \vect{C}_{K_t k}
\end{IEEEeqnarraybox*} \right] \quad \textrm{where} \quad \vect{C}_{jk} = \begin{cases}
\vect{I}_{N_j}, & \textrm{if }  k \in \mathcal{C}_j,\\[-1.2ex]
\vect{0}_{N_j}, & \textrm{otherwise}.
\end{cases}
\end{equation}
Thus, $\vect{h}_{kc}^H \vect{D}_k $ is the channel that carries data to $\textrm{MS}_k$ and $\vect{h}_{kc}^H \vect{C}_k $ is the channel that carries
non-negligible interference. If the received signal by $\textrm{MS}_k$ at subcarrier $c$ is denoted  $y_{kc} \in \mathbb{C}$, we have
\begin{equation} \label{eq_system_model}
y_{kc}= \vect{h}_{kc}^H \vect{C}_k \sum_{\bar{k}=1}^{K_r}
\vect{D}_{\bar{k}} \vect{s}_{\bar{k}c} + n_{kc}
\end{equation}
where $\vect{s}_{kc} \in \mathbb{C}^{N \times 1}$ is the data symbol vector to $\textrm{MS}_k$. This random vector is zero-mean and has
signal correlation matrix $\vect{S}_{kc} = \mathbb{E}\{ \vect{s}_{kc} \vect{s}_{kc}^H \}$. In essence, resource allocation means selection of $\vect{S}_{kc}$.
The special case of $\rank(\vect{S}_{kc})=1$ is known as single-stream beamforming and is particulary simple to implement \cite{Goldsmith2003a}, but for now we let $\vect{S}_{kc}$ have arbitrary rank in the performance optimization.

The additive zero-mean term $n_{kc}$ has variance $\sigma_{kc}^2$
and it models both noise and weak interference from the transmitters
with $k \not \in \mathcal{C}_j$ (see Definition \ref{def_dynamic_clusters}). This assumption limits the
amount of CSI required to model the transmission and is reasonable
if transmitters coordinate interference to all cell edge
terminals of adjacent cells. In the analysis, $\textrm{BS}_j$ is
assumed to know the channels $\vect{h}_{jkc}$ perfectly to all
$\textrm{MS}_k$ with $k \in \mathcal{C}_j$.

In \eqref{eq_system_model}, perfect synchronization is assumed between transmitters that jointly
serve a terminal with data (synchronization uncertainty is
considered in Section \ref{section_numerical_examples}).
Joint transmissions are also assumed to create synchronous interference \cite{Zhang2008a};
this is reasonable for coordination between adjacent cells, while larger scales would require unacceptably long cyclic prefixes.

The transmission is limited by $L$ linear power constraints,
\begin{equation} \label{eq_power_constraint}
\sum_{k=1}^{K_r} \sum_{c=1}^{K_c} \tr\{ \vect{Q}_l \vect{S}_{kc}\} \leq q_l,
\end{equation}
represented by matrices $\vect{Q}_l \succeq \vect{0}_N$. To ensure that the total power is constrained and only is allocated to dimensions used for transmission, these matrices must satisfy two conditions: a) $\vect{Q}_{lk} - \vect{D}_{k}^H \vect{Q}_{lk} \vect{D}_{k}$ is diagonal and b) $\sum_{l=1}^{L} \vect{Q}_{lk} \succ \vect{0}_N \, \forall k$.
These constraints can, for example, be defined on the total transmit power (most power efficient), per-transmitter power (controls the radiated power in an area), or per-antenna power (protects the dynamic range of each power amplifier).
It is straightforward also include subcarrier specific constraints (e.g., frequency masks), but that would complicate the notation.

\subsection{Examples: Two Simple Multicell Scenarios}

The purpose of the proposed dynamic cooperation clusters is to jointly describe and analyze a variety of multicell scenarios. Typical examples are ideal network MIMO \cite{Karakayali2006a} (where all transmitters serve all terminals) and interference channels \cite{Shang2010a} (with only one unique terminal per transmitter):

\subsubsection{Ideal Network MIMO} All transmitters serve and coordinate interference to all terminals, meaning that $\vect{D}_k=\vect{C}_k=\vect{I}_N$ for all $k$. If a total power constraint is used, then $L=1$ and $\vect{Q}_1=\vect{I}_N$. If per-antenna constraint are used, then $L=N$ and $\vect{Q}_l$ is only non-zero at the $l$th diagonal element.

\subsubsection{Two-user Interference Channel} Let $\textrm{BS}_k$ serve $\textrm{MS}_k$ and coordinate interference to the other receiver. Then, $\vect{D}_1=\left[\begin{IEEEeqnarraybox*}[\mysmallarraydecl]
[c]{,c,c,}
\vect{I}_{N_1}& \vect{0}\\
\vect{0}& \vect{0}%
\end{IEEEeqnarraybox*}\right]$ and $\vect{D}_2=\left[\begin{IEEEeqnarraybox*}[\mysmallarraydecl]
[c]{,c,c,}
\vect{0} & \vect{0}\\
\vect{0}& \vect{I}_{N_2}%
\end{IEEEeqnarraybox*}\right]$, while $\vect{C}_1=\vect{C}_2=\vect{I}_N$. If each transmitter has its own total power constraint, then $L=2$ and $\vect{Q}_l=\vect{D}_l$ for $l=1,2$.

\section{Problem Formulation}
\label{section_problem_formulation}

In this section, we will define two optimization problems. The main
problem is to \emph{maximize the system performance with arbitrary
monotonic utility functions} \eqref{eq_P1}, which in general is a
non-convex problem. For this reason, we also consider the
problem to \emph{maximize performance with individual quality
constraints} \eqref{eq_P2}, which is convex
and thus efficiently solved using
general-purpose optimization algorithms \cite{Boyd2004a} and solvers \cite{cvx}. In both cases, we make the assumption of
single-user detection (SUD) \cite{Shang2010a}, which means that
receivers treat co-terminal interference as noise (i.e., not
attempting to decode and subtract interference). This assumption
leads to suboptimal performance, but is important to achieve simple
and practical receivers. Under SUD, the signal-to-interference-and-noise ratio
(SINR) for $\textrm{MS}_k$ at subcarrier $c$ becomes
\begin{equation} \label{eq_SINR_DL_general}
\begin{split}
\textrm{SINR}^{\textrm{DL}}_{kc} (\vect{S}_{1
c},&\ldots,\vect{S}_{K_r c}) \\
& =  \frac{
\vect{h}_{kc}^H \vect{C}_k \vect{D}_k \vect{S}_{kc} \vect{D}_k^H \vect{C}_k^H \vect{h}_{kc}
}{\sigma_{kc}^2 \!+\! \vect{h}_{kc}^H \vect{C}_{k} (
\fracSum{\bar{k} \neq k} \vect{D}_{\bar{k}} \vect{S}_{\bar{k}c}
\vect{D}_{\bar{k}}^H ) \vect{C}_{k}^H \vect{h}_{kc}} \\ & =  \frac{
\vect{h}_{kc}^H \vect{D}_k \vect{S}_{kc} \vect{D}_k^H \vect{h}_{kc}
}{\sigma_{kc}^2 \!+\! \vect{h}_{kc}^H \vect{C}_{k} (
\fracSum{\bar{k} \in \mathcal{I}_k} \vect{D}_{\bar{k}}
\vect{S}_{\bar{k}c} \vect{D}_{\bar{k}}^H ) \vect{C}_{k}^H
\vect{h}_{kc}}
\end{split}
\end{equation}
where the last equality follows from $\vect{C}_{k}
\vect{D}_{k}=\vect{D}_{k}$ and that $\vect{C}_{k}
\vect{D}_{\bar{k}} \neq \vect{0}$ only for terminals $\bar{k}$ in
\begin{equation}
\mathcal{I}_k = \bigcup_{
\{j \in \mathcal{J}; \, k \in \mathcal{C}_j\}
}
\mathcal{D}_j \! \setminus \! \{ k \}.
\end{equation}
This set contains all co-terminals that are served by the transmitters that serve $\textrm{MS}_k$.

The achievable performance of a terminal can be measured in
different ways, but the most common quality measures are all
monotonic functions of the SINR: data rate, mean square
error (MSE), and bit/symbol error rate (BER/SER)
\cite{Palomar2003a}. Herein, we describe the
performance of $\textrm{MS}_k$ by an arbitrary quality function
$g_k(\cdot)$ of the SINRs that is \emph{strictly monotonically
increasing}\footnote{We use the convention that the performance
measure is a function to be maximized. Thus, if the problem is to
minimize the error (e.g., MSE, BER, or SER), we maximize the error
with a negative sign.} in each argument:
$g_{k}(\textrm{SINR}^{\textrm{DL}}_{k1},\ldots,\textrm{SINR}^{\textrm{DL}}_{k
K_c})$. The system performance represents a balance between
maximizing the performance achieved by the different terminals.
Herein, we represent it by an arbitrary system utility function $f(\cdot)$ of the
terminal performances that is also strictly monotonically increasing. We arrive at the following general optimization problem:
\begin{equation} \label{eq_P1} \tag{P1}
\begin{aligned}
\maximize{\{ \vect{S}_{kc} \}_{k=1,c=1}^{K_r,K_c} }\,\, & f \left(
g_{1},\ldots,g_{K_r}\right) \\
\mathrm{subject}\,\,\mathrm{to}\,\,\, & \, \, \vect{S}_{kc} \succeq \vect{0} \,\,\,\, \forall k,c, \\
g_{k} = g_{k} & \left(
\left\{ \textrm{SINR}^{\textrm{DL}}_{kc} (\vect{S}_{1
c},\ldots,\vect{S}_{K_r c}) \right\}_{c=1}^{K_c} \right) \,\,\,\, \forall k, \\
& \, \sum_{k=1}^{K_r} \sum_{c=1}^{K_c} \tr\{ \vect{Q}_l \vect{S}_{kc}\} \leq q_l \,\,\,\, \forall l.
\end{aligned}
\end{equation}
The system utility function $f(\cdot)$ can for example represent the weighted
max-min terminal performance $f( g_{1},\ldots,g_{K_r})=\min_{k}  g_k/\mu_k$
or the weighted sum performance $f( g_{1},\ldots,g_{K_r})=\sum_k
\mu_k g_k$ for some collection of weights $\mu_k \geq 0$. Although
many other functions are possible, it is worth noting that all reasonable
performance optimizations can be expressed in terms of these utility functions.\footnote{Consider the region of achievable
performance points $(g_1,\ldots,g_{K_r})$. If this region is convex, there exists a supporting hyperplane
$\sum_k \mu_k g_k = \mathrm{constant}$ for each point on the outer
boundary \cite[Theorem 1.5]{Tuy1998a}. The weights $\mu_k$ of this hyperplane defines a weighted sum
performance optimization that ends up in that point. Even if the region is non-convex, a line can be
drawn from the origin to any point on the outer boundary and the direction of this line defines a weighted
max-min terminal performance optimization that ends up in that boundary point.} Observe that
\eqref{eq_P1} solves the whole resource allocation (i.e., precoding and scheduling over subcarriers) as a single optimization problem. Thus, it is not unexpected that \eqref{eq_P1} is non-convex and generally NP-hard \cite{Liu2011a}, making it very difficult to design numerical search algorithms with global convergence.

The second optimization problem is designed to be easier to solve.
It is based on satisfying predefined quality of service (QoS)
constraints $\{ \gamma_{kc} \}$; that is, $\textrm{SINR}^{\textrm{DL}}_{kc} \geq \gamma_{kc}$ for each terminal $k$ and subcarrier $c$. We pose it as the following feasibility problem:
\begin{equation} \label{eq_P2} \tag{P2}
\begin{aligned}
\mathrm{find} \,\, & \,\, \{ \vect{S}_{kc} \}_{k=1,c=1}^{K_r,K_c} \\
\mathrm{subject}\,\,\mathrm{to}\,\, & \,\,  \textrm{SINR}^{\textrm{DL}}_{kc} (\vect{S}_{1
c},\ldots,\vect{S}_{K_r c}) \geq \gamma_{kc}\,\,\, \forall k,c, \\
& \, \, \sum_{k=1}^{K_r} \sum_{c=1}^{K_c} \tr\{ \vect{Q}_l \vect{S}_{kc}\} \leq q_l \,\,\,\, \forall l, \\
& \, \, \vect{S}_{kc} \succeq \vect{0} \,\,\,\, \forall k,c.
\end{aligned}
\end{equation}
This problem is similar to the convex single-carrier problems in \cite{Boche2002a,Wiesel2006a,Yu2007a,Dahrouj2010a}, but a main difference from \eqref{eq_P2} is that they multiply each $q_l$ with a parameter $\beta$ and minimize over it. Such power minimization might lead to $\beta>1$, which corresponds to breaching some of the power constraints. Fortunately, the multi-carrier problem in \eqref{eq_P2} with fixed power constraints is also convex:
\begin{lemma}
The constraints in \eqref{eq_P2} can be formulated as semi-definite constraints and thus \eqref{eq_P2} is convex.
\end{lemma}
\begin{IEEEproof}
The power constraints are already semi-definite and the QoS constraints can be reformulated in a semi-definite way as
\begin{equation}
\begin{split}
\fracSum{\bar{k} \in \mathcal{I}_k} \tr\{ \gamma_{kc} \vect{D}_{\bar{k}}^H & \vect{C}_{k}^H
\vect{h}_{kc} \vect{h}_{kc}^H \vect{C}_{k}
 \vect{D}_{\bar{k}}
\vect{S}_{\bar{k}c}  \} \\
&- \tr\{ \vect{D}_k^H \vect{h}_{kc} \vect{h}_{kc}^H \vect{D}_k \vect{S}_{kc} \}  \leq -\gamma_{kc} \sigma_{kc}^2
\end{split}
\end{equation}
by moving the denominator of $\textrm{SINR}^{\textrm{DL}}_{kc}$ to the right hand side of the constraint and exploiting that $\vect{a}^H \vect{b} = \tr\{ \vect{a}^H \vect{b}\} = \tr\{ \vect{b} \vect{a}^H \}$ for any compatible vectors $\vect{a}$ and $\vect{b}$.
\end{IEEEproof}

The convexity makes it easier to analyze and solve \eqref{eq_P2}. In Section \ref{section_optimality_properties}, \eqref{eq_P1} and \eqref{eq_P2} are analyzed and common properties of their solutions are derived. The joint analysis is based on the following relationship, which is easily proved by contradiction:

\begin{lemma} \label{lemma_relation_P1_P2}
If the optimal SINRs, $\textrm{SINR}^{*\textrm{DL}}_{kc}$, for each terminal $k$ and subcarrier $c$ in \eqref{eq_P1}
are used as QoS constraints $\gamma_{kc}$ in \eqref{eq_P2}, then all
solutions to \eqref{eq_P2} are also optimal for \eqref{eq_P1}.
\end{lemma}

Thus, any property of \eqref{eq_P2} that holds for any set of $\{\gamma_{kc}\}$ also holds for \eqref{eq_P1}. The price of achieving the convex problem formulation in
\eqref{eq_P2} is that the system must propose the individual QoS
constraints. In general the optimal QoS values of \eqref{eq_P1} are unknown, but the solution to
\eqref{eq_P1} can in theory be achieved by iteratively solving \eqref{eq_P2} for different QoS constraints.
Unfortunately, the available algorithms either have slow convergence as in \cite{Brehmer2010a,Bjornson2012a}
or cannot guarantee global convergence as in \cite{Zheng2008a,Tolli2008a,Venturino2010a}. This motivates the search for
properties of the optimal solutions that can simplify the optimization or be used to achieve efficient suboptimal solutions.

\section{Properties of Optimal Resource Allocation}
\label{section_optimality_properties}

In this section, we will derive three properties of the optimal solutions to \eqref{eq_P1} and \eqref{eq_P2}:
\begin{itemize}
\item Optimality of single-stream beamforming;
\item Conditions for full power usage;
\item Parametrization of optimal precoding strategies.
\end{itemize}
Taking these optimality properties into account when solving \eqref{eq_P1} will greatly reduce the search space for optimal solutions. They are used in Section \ref{section_distributed_precoding} to achieve low-complexity solutions that perform very well in the measurement-based evaluation in Section \ref{section_numerical_examples}.

\subsection{Optimality of Single-Stream Beamforming}

The first optimality property of \eqref{eq_P1} and \eqref{eq_P2} is the sufficiency of considering
signal correlation matrices $\vect{S}_{kc}$ that are rank one. This might seem intuitive and is
often assumed for single-antenna receivers without discussion (see \cite{Yu2007a,Dahrouj2010a,Tolli2008a,Zheng2008a,Tolli2009b,Venturino2010a}).
However, the following toy example shows that it is actually not a necessary condition under general utility functions and power constraints
(i.e., high-rank solutions can give the same performance, but never better, than the rank-one solutions):

\begin{example}
Consider a single-carrier point-to-point system ($K_t=K_r=K_c=1$) with two transmit antennas, the channel vector $\vect{h}_{11}=[1\,\, 0]^T$, and the per-antenna power constraints
$\tr\left\{  \left[\begin{IEEEeqnarraybox*}[\mysmallarraydecl]
[c]{,c,c,}
1& 0\\
0& 0%
\end{IEEEeqnarraybox*}\right] \vect{S}_{11} \right\} \leq 1$ and $\tr\left\{  \left[\begin{IEEEeqnarraybox*}[\mysmallarraydecl]
[c]{,c,c,}
0& 0\\
0& 1%
\end{IEEEeqnarraybox*}\right] \vect{S}_{11} \right\} \leq 1$.
Under these conditions, \eqref{eq_P1} is solved optimally by both the rank-one single correlation matrix $\vect{S}_{11}=\left[\begin{IEEEeqnarraybox*}[\mysmallarraydecl]
[c]{,c,c,}
1& 0\\
0& 0%
\end{IEEEeqnarraybox*}\right]$ and by the rank-two matrix $\vect{S}_{11}=\left[\begin{IEEEeqnarraybox*}[\mysmallarraydecl]
[c]{,c,c,}
1& 0\\
0& 1%
\end{IEEEeqnarraybox*}\right]$.
\end{example}

To prove the sufficiency of rank-one signal correlation matrices, we start with a lemma from \cite{Wiesel2008a}.

\begin{lemma} \label{lemma_rank_one}
The convex optimization problem
\begin{equation}
  \tag{P3}
  \label{eq_P3}
\begin{aligned}
\maximize{ \vect{X} \succeq 0}\,\, & \,\, \vect{a}^H \vect{X} \vect{a}\\
\mathrm{subject}\,\,\mathrm{to}\,\, & \,\, \tr\{\vect{B}_i \vect{X}\} \leq b_i \quad \textrm{for } i=1,\ldots,M,
\end{aligned}
\end{equation}
with arbitrary Hermitian matrices $\vect{B}_i \succeq 0$ and scalars
$b_i \geq 0 \,\, \forall i$, has solutions with $\rank(\vect{X})\leq 1$.
\end{lemma}
\begin{IEEEproof}
The proof is given in \cite[Appendix III]{Wiesel2008a}.
\end{IEEEproof}

The main result in this subsection is the following theorem.

\begin{theorem} \label{theorem_beamforming_optimality}
It holds for \eqref{eq_P1} and \eqref{eq_P2} that
\begin{itemize}
\item If the problem is feasible, there exists single-stream beamforming solutions (i.e., $\rank(\vect{S}_{kc})\leq 1 \,\, \forall k,c$).
\end{itemize}
\end{theorem}
\begin{IEEEproof}
Let $\{\vect{S}^*_{kc}\}_{k=1, c=1}^{K_r, K_c}$ be an optimal
solution to \eqref{eq_P1}. For each such optimal signal correlation matrix, we can create the problem
\begin{equation} \label{eq_function_class}
\begin{split}
\maximize{\vect{S}_{kc} \succeq \vect{0}}\,\, & \quad \vect{h}_{kc}^H \vect{D}_k
\vect{S}_{kc} \vect{D}_k^H \vect{h}_{kc}
\\
\mathrm{subject}\,\,\mathrm{to}\,\, & \quad \vect{h}_{\bar{k}c}^H
\vect{C}_{\bar{k}} \vect{D}_k \vect{S}_{kc} \vect{D}_k^H \vect{C}_{\bar{k}}^H \vect{h}_{\bar{k}c} \leq z_{k \bar{k} c}^2 \,\, \forall \bar{k} \neq k, \\
 & \quad \tr\{ \vect{Q}_{l} \vect{S}_{kc}  \} \leq
q_{lkc} \,\,\, \forall l,
\end{split}
\end{equation}
where $z_{k \bar{k} c}^2=\vect{h}_{\bar{k}c}^H \vect{C}_{\bar{k}}
\vect{D}_k \vect{S}^*_{kc} \vect{D}_k^H \vect{C}_{\bar{k}}^H
\vect{h}_{\bar{k}c}$ and  $q_{lkc} = \tr\{ \vect{Q}_{l} \vect{S}^*_{kc} \}$. This problem tries to maximize the signal gain under the constraint that neither more interference is caused nor more transmit power is used than with $\vect{S}^*_{kc}$. Obviously, $\vect{S}^*_{kc}$ is an optimal solution to \eqref{eq_function_class} (if
better solutions would have existed, these could have be used to improve the utility in \eqref{eq_P1}, which is a contradiction).

Now, observe that \eqref{eq_function_class} has the structure of \eqref{eq_P3} in Lemma \ref{lemma_rank_one} (since $\vect{h}_{\bar{k}c}^H
\vect{C}_{\bar{k}} \vect{D}_k \vect{S}_{kc} \vect{D}_k^H \vect{C}_{\bar{k}}^H \vect{h}_{\bar{k}c}= \tr\{ \vect{D}_k^H \vect{C}_{\bar{k}}^H \vect{h}_{\bar{k}c} \vect{h}_{\bar{k}c}^H
\vect{C}_{\bar{k}} \vect{D}_k \vect{S}_{kc}\}$). Thus, if $\vect{S}^*_{kc}$ has rank greater than one, Lemma \ref{lemma_rank_one} shows that there exist an alternative solution
$\vect{S}^{**}_{kc}$ with $\rank(\vect{S}^{**}_{kc})\leq 1$. This solution can be used instead of $\vect{S}^{*}_{kc}$ without decreasing the performance. The proof for \eqref{eq_P2} follows along the same lines.
\end{IEEEproof}

The optimality of single-stream beamforming both decreases the search space for optimal solutions
and makes the solution easier to implement (since vector coding or successive interference
cancelation is required if $\rank(\vect{S}_{kc})>1$ \cite{Goldsmith2003a}). Observe
that $\rank(\vect{S}_{kc})\leq 1$ in Theorem \ref{theorem_beamforming_optimality} implies
that the rank might be zero, which corresponds to $\vect{S}_{kc}=\vect{0}$ (i.e., no transmission to $\textrm{MS}_k$ at subcarrier $c$).

Recently, similar optimality results for single-stream beamforming have been derived for a few special multicell scenarios. The MISO interference channel was considered in \cite{Shang2010a} and a certain class of multicell systems was considered in \cite{Mochaourab2011a}. Per-transmitter power constraints were considered in both \cite{Shang2010a} and \cite{Mochaourab2011a}, making Theorem \ref{theorem_beamforming_optimality} a generalization to arbitrary linear power constraints and our general multicell framework.

\subsection{Conditions for Full Power Usage}

If only the total power usage over all transmitters is constrained,
it is trivial to prove that the solution to \eqref{eq_P1} will use all available power.
Under general linear power constraints, it may be better to not use full
power at each transmitter or antenna; there is a balance between increasing signal gains and limiting co-terminal interference.
This is illustrated by the following toy example:

\begin{example}
Consider a two-user interference channel with single-antenna transmitters ($K_t=K_r=2$, $N_1=N_2=K_c=1$) and the channel vectors $\vect{h}_{11}=[1 \,\, \sqrt{1/10}]^T$ and $\vect{h}_{21}=[\sqrt{1/2} \,\, 1]^T$. $\textrm{BS}_j$ transmits to $\textrm{MS}_j$ and coordinates interference to both terminals, meaning that $\vect{D}_1=\left[\begin{IEEEeqnarraybox*}[\mysmallarraydecl]
[c]{,c,c,}
1& 0\\
0& 0%
\end{IEEEeqnarraybox*}\right]$, $\vect{D}_2=\left[\begin{IEEEeqnarraybox*}[\mysmallarraydecl]
[c]{,c,c,}
0& 0\\
0& 1%
\end{IEEEeqnarraybox*}\right]$, and $\vect{C}_1=\vect{C}_2=\vect{I}_2$. The per-transmitter power is constrained as $\tr\left\{ \vect{D}_j \vect{S}_{j1} \right\} \leq 20 \,\,\, \forall j$.

Under max-min rate optimization with $f(\textrm{SINR}_1,\textrm{SINR}_2)=\min_k \log_2(1+\textrm{SINR}_k)$ and identical quality functions, the optimal solution to \eqref{eq_P1} is $\vect{S}_{11}=10 \vect{D}_1$ and $\vect{S}_{21}=20 \vect{D}_2$. This solution gives $\textrm{SINR}_1=\textrm{SINR}_2=10/3$, and observe that only $\textrm{BS}_2$ uses full power. If $\textrm{BS}_1$ would increase its power usage, then $\textrm{SINR}_2$ would decrease and thus the performance would be degraded.
\end{example}

In principle, the knowledge that a certain power constraint is active removes one dimension from the optimization problem. The second optimality property provides conditions on when full power should be used in general multicell systems.

\begin{theorem} \label{theorem_power_usage}
It holds for \eqref{eq_P1} and \eqref{eq_P2} that
\begin{itemize}
\item There exist solutions that satisfy at least one power constraint with equality.

\item If there are only per-transmitter power constraints, $\textrm{BS}_j$ should use full power if $|\mathcal{C}_j|\leq N_j$.
\end{itemize}
\end{theorem}
\begin{IEEEproof}
For a given optimal solution $\{\vect{S}^*_{kc}\}_{k=1, c=1}^{K_r, K_c}$, assume that all power constraints
in \eqref{eq_power_constraint} are inactive. Let $\varepsilon=\max_l q_l/( \sum_{k=1}^{K_r} \sum_{c=1}^{K_c} \tr\{ \vect{Q}_l \vect{S}^*_{kc}\})$
and observe that $\varepsilon>1$. Then, $\{ \varepsilon \vect{S}^*_{kc}\}_{k=1, c=1}^{K_r, K_c}$ will satisfy all power constraints and at least one
of them becomes active. The performance is not decreased since the factor $\varepsilon$ can be seen as decreasing the relative noise
power in each SINR in \eqref{eq_SINR_DL_general}. Thus, there always exist a solution that satisfies at least one power constraint with equality.

The second part requires that there exists a $k \in \mathcal{D}_j$ with $\vect{h}_{jkc} \not \in
\mathrm{span}( \bigcup_{\bar{k} \in \mathcal{C}_j \setminus \{k\}}
\{ \vect{h}_{j\bar{k}c}\} )$. If $|\mathcal{C}_j|\leq N_j$, this is satisfied with probability one for stochastic channels (with non-singular
covariance matrices). Then, it exists a zero-forcing beamforming vector that can increase the signal gain at $\textrm{MS}_k$
without causing interference to any other terminals in $\mathcal{C}_j \setminus \{k\}$. By increasing the power in
the zero-forcing direction until the power constraint for transmitter $j$ is active, the second part of the theorem
is proved. The details are along the lines of \cite[Theorem 2]{Bjornson2010c}.
\end{IEEEproof}

The interpretation is that at least one power constraint should be active in the optimal solution. In addition, the fewer terminals that a transmitter coordinates interference to, the more power can it use. The second item in Theorem \ref{theorem_power_usage} can be relaxed to that full power is required at $\textrm{BS}_j$ if fewer than $N_j$ terminals in $\mathcal{C}_j$ are allocated to some subcarrier $c$.

\subsection{Parametrization of Optimal Precoding Strategies}

The third optimality property is an explicit parametrization of the optimal solution to \eqref{eq_P1} using $K_r K_c+L$ parameters
between zero and one. In this context, \emph{explicit} means that the parameters give a transmit strategy directly, without having to solve any optimization problem.\footnote{
Fewer parameters can be achieved by using the $K_r K_c$ QoS constraints $\{\gamma_{kc}\}$ in \eqref{eq_P2} as parameters (see \cite{Bjornson2012a}), but this is impractical since
finding the corresponding transmit strategy means solving a convex optimization problem.} Recalling that \eqref{eq_P1} consists
of finding $K_r K_c$ complex-valued positive semi-definite transmit correlation matrices of size $N \times N$,
this constitutes a major reduction of the search space for the optimal solution.

As a first step, we exploit the single-stream beamforming optimality in Theorem \ref{theorem_beamforming_optimality} to derive a dual to
the feasibility problem \eqref{eq_P2}. The following lemma builds upon the line of work in \cite{Boche2002a,Wiesel2006a,Yu2007a,Dahrouj2010a}
and characterizes the solution to
\eqref{eq_P2} through the principle of virtual uplink-downlink duality. To simplify the notation, we first define
$\widetilde{\mathcal{I}}_k$ as the set of terminals that transmitters
serving $\textrm{MS}_k$ are coordinating interference to. Formally,
\begin{equation}
\widetilde{\mathcal{I}}_k = \bigcup_{
\{j \in \mathcal{J}; \, k \in \mathcal{D}_j\}
} \mathcal{C}_j \! \setminus \! \{ k \}.
\end{equation}

\begin{lemma} \label{lemma_duality}
Strong duality holds between \eqref{eq_P2} and the Lagrange dual problem
\begin{align} \label{eq_dual_to_P2} \tag{D2}
\maximize{\boldsymbol{\omega}, \{ \boldsymbol{\lambda}_c\}_{c=1}^{K_c}}\,\, & \,\,
\sum_{k=1}^{K_r} \sum_{c=1}^{K_c} \lambda_{kc}
\sigma_{kc}^2  -
\sum_{l=1}^{L} \omega_l q_l \\
\mathrm{subject}\,\,\mathrm{to}\,\, & \,\,  \omega_l \geq 0, \, \lambda_{kc} \geq
0 \quad \forall k,c,l, \notag \\
\max_{\bar{\vect{w}}_{kc}} \, &
\textrm{SINR}^{\textrm{VUL}}_{kc}(\bar{\vect{w}}_{1
c},\ldots,\bar{\vect{w}}_{K_r c}, \boldsymbol{\omega}, \boldsymbol{\lambda}_c)
= \gamma_{kc}, \,\,\, \forall k,c, \notag
\end{align}
with $\boldsymbol{\omega}=[\omega_1,\ldots,\omega_{L}]^T$,
$\boldsymbol{\lambda}_c=[\lambda_{1c},\ldots,\lambda_{K_r c}]^T$, and
\begin{equation} \label{eq_SINR_UL_beamforming}
\begin{split}
&\textrm{SINR}^{\textrm{VUL}}_{kc} (\bar{\vect{w}}_{1
c},\ldots,\bar{\vect{w}}_{K_r c}, \boldsymbol{\omega}, \boldsymbol{\lambda}_c) \\ &=
\frac{ \lambda_{kc} \bar{\vect{w}}_{kc}^H
\vect{D}_k^H \vect{h}_{kc} \vect{h}_{kc}^H \vect{D}_k
\bar{\vect{w}}_{kc}}{ \bar{\vect{w}}_{kc}^H (
\fracSum{l}  \omega_l \vect{Q}_l \!+\! \fracSum{\bar{k} \in
\widetilde{\mathcal{I}}_k} \lambda_{\bar{k}c} \vect{D}_k^H \vect{C}_{\bar{k}}^H
\vect{h}_{\bar{k}c} \vect{h}_{\bar{k}c}^H \vect{C}_{\bar{k}} \vect{D}_k )
\bar{\vect{w}}_{kc}}.
\end{split}
\end{equation}
The optimal utility of \eqref{eq_dual_to_P2} satisfies $\sum_{k=1}^{K_r} \sum_{c=1}^{K_c} \lambda_{kc}
\sigma_{kc}^2  - \sum_{l=1}^{L} \omega_l q_l=0$  and the optimal $\vect{S}_{kc}$ in \eqref{eq_P2} is equal to
the optimal $\bar{\vect{w}}_{kc} \bar{\vect{w}}_{kc}^H$ up to a scaling factor.
\end{lemma}
\begin{IEEEproof}
The proof is given in Appendix \ref{appendix_proof_lemma_duality}.
\end{IEEEproof}

The main result in this subsection is the following theorem that exploits Lemma \ref{lemma_duality} to derive an explicit parametrization for the solution to \eqref{eq_P1}.

\begin{theorem} \label{theorem_beamforming_parametrization}
The optimal solution to \eqref{eq_P1} can be expressed as $\vect{S}_{kc} \!=\! p_{kc} \vect{v}_{kc} \vect{v}_{kc}^H$.
For some choice of $\omega_{l},\lambda_{kc} \in [0,1]$
($l=1,\ldots,L$, $k=1,\ldots,K_r$, $c=1,\ldots,K_c$),
the optimal power allocation $p_{kc}$ and beamforming vectors $\vect{v}_{kc}$ are given by
\begin{equation} \label{eq_parameterization_beamformer}
\vect{v}_{kc} \!=\!
\frac{ \Big( \! \sum_{l=1}^{L}  \omega_l \vect{Q}_l  \!+\!\!
\fracSum{\bar{k} \in \widetilde{\mathcal{I}}_k} \lambda_{\bar{k}c}
\vect{D}_k^H \vect{C}_{\bar{k}}^H \vect{h}_{\bar{k}c}
\vect{h}_{\bar{k}c}^H \vect{C}_{\bar{k}} \vect{D}_k \!\Big)^{\!\dagger}
\vect{D}_k^H \vect{h}_{kc} }{\Big\| \!\Big( \!\sum_{l=1}^{L}  \omega_l \vect{Q}_l  \!+\!\!
\fracSum{\bar{k} \in \widetilde{\mathcal{I}}_k} \lambda_{\bar{k}c}
\vect{D}_k^H \vect{C}_{\bar{k}}^H \vect{h}_{\bar{k}c}
\vect{h}_{\bar{k}c}^H \vect{C}_{\bar{k}} \vect{D}_k \! \Big)^{\!\dagger}
\vect{D}_k^H \vect{h}_{kc}  \Big\| }
\end{equation}
for all $k,c$ and
\begin{equation} \label{eq_parameterization_powerallocation}
\left[\begin{IEEEeqnarraybox*}[][ccc]{ccc} p_{1c} \, &
\ldots &
\,p_{K_r c}%
\end{IEEEeqnarraybox*}\right] =
\left[\begin{IEEEeqnarraybox*}[][ccc]{ccc} \gamma_{1 c} \sigma_{1 c}^2 \, &
\ldots &
 \, \gamma_{K_r c} \sigma_{K_r c}^2%
\end{IEEEeqnarraybox*}\right]
\vect{M}_c^{\dagger} \quad \forall c.
\end{equation}
Here, the $mn$th element of $\vect{M}_c \in \mathbb{R}^{K_r \times K_r}$ is
\begin{equation} \label{eq_parameterization_M-matrix}
[ \vect{M}_c ]_{mn} = \begin{cases} | \vect{h}_{mc}^H \vect{D}_m
\vect{v}_{mc} |^2, & m = n, \\ - \gamma_{nc} | \vect{h}_{nc}^H \vect{C}_{n}
\vect{D}_{m} \vect{v}_{mc}|^2, & m \neq n, \end{cases}
\end{equation}
and
\begin{equation}
\begin{split}
\gamma_{kc} = \lambda_{kc} & \vect{h}_{kc}^H \vect{D}_k
\Big( \sum_{l}  \omega_l\vect{Q}_l \\ & + \fracSum{\bar{k} \in
\widetilde{\mathcal{I}}_k} \lambda_{\bar{k}c}
 \vect{D}_k^H \vect{C}_{\bar{k}}^H \vect{h}_{\bar{k}c} \vect{h}_{\bar{k}c}^H \vect{C}_{\bar{k}} \vect{D}_k
\Big)^{\dagger} \vect{D}_k^H \vect{h}_{kc}.
\end{split}
\end{equation}
\end{theorem}
\begin{IEEEproof}
Based on Lemma \ref{lemma_relation_P1_P2}, there exists a set of $\{ \gamma_{kc} \}$ such that the solutions to \eqref{eq_P2} are solutions to \eqref{eq_P1}. Using these $\{ \gamma_{kc} \}$,
we apply Lemma \ref{lemma_duality} to achieve $\vect{v}_k$ as the normalized solution to the generalized eigenvalue problem in \eqref{eq_dual_to_P2}.
To determine $p_{kc}$ for all $k$ and $c$, observe that since we consider the optimal solution to \eqref{eq_P1}, all QoS constraints in \eqref{eq_P2} needs to be satisfied with equality.
These constraints gives $K_r K_c$ linear equations that can be expressed and
solved as in \eqref{eq_parameterization_powerallocation}. The
alternative expression for $\gamma_{kc}$ in \eqref{eq_parameterization_M-matrix} is achieved from \eqref{eq_rewrite_with_receive_beamformer} in Appendix \ref{appendix_proof_lemma_duality}.

It remains to show that it is sufficient to search for Lagrange
multipliers $\omega_{l},\lambda_{kc} \in [0,1]$. Dual feasibility in \eqref{eq_dual_to_P2} requires $\omega_{l},\lambda_{kc} \geq 0$. Observe that \eqref{eq_parameterization_beamformer} and
\eqref{eq_parameterization_powerallocation} are unaffected by a
common scaling factor in $\omega_{l},\lambda_{kc}$. Thus, if any of the variables is greater than one, we can divide all the variables with $d = \max(\max_l
\omega_{l}, \max_{k,c} \lambda_{kc})$ and achieve a set of parameters between zero and one. Thus, it is sufficient with $\omega_{l},\lambda_{kc} \in [0,1]$.
\end{IEEEproof}

This theorem shows that the whole resource allocation in \eqref{eq_P1} (i.e, precoding and scheduling over subcarriers) is governed by $K_r K_c +L$
real-valued parameters, each between zero and one. Thus, even if multiple base stations are involved in the transmission, there is
only a single parameter per terminal and subcarrier. Remarkably, this simple structure holds for any outer utility
function $f(\cdot)$ and terminal quality functions $g_k(\cdot)$ that
are strictly monotonically increasing.

The parameter $\lambda_{kc}$ in Theorem \ref{theorem_beamforming_parametrization} represents the priority of $\textrm{MS}_k$ at subcarrier $c$. This is easily understood from
\eqref{eq_parameterization_M-matrix} by observing that
\begin{equation} \label{eq_impact_of_lambda}
\frac{\partial}{\partial \lambda_{kc}} \textrm{SINR}^{\textrm{DL}}_{kc}=\frac{\partial}{\partial \lambda_{kc}} \gamma_{kc}
\begin{cases} \geq 0, & k = \bar{k}, \\ \leq 0, & k \neq \bar{k}.
\end{cases}
\end{equation}
Clearly, $\lambda_{kc}>0$ only for terminals allocated to subcarrier $c$. In addition, all SINRs are decreasing functions of $\omega_{l}$, and it is worth noting
that $\omega_{l}=0$ for all inactive power constraints.

Similar parametrizations have been derived for single-carrier systems in \cite{Jorswieck2008b,Zhang2010a,Shang2010a, Mochaourab2011a,Bjornson2010c}.
For $K_r$-user MISO interference channels, a characterization using $K_t(K_r-1)$ complex-valued parameters was
derived in \cite{Jorswieck2008b}. It was improved in \cite{Zhang2010a} by making them positive real-valued, and even earlier in \cite{Shang2010a} by using
$K_t(K_r-1)$ parameters from $[0,1]$.\footnote{From a complexity perspective, there is little difference between parameters from $[0,1]$ and $[0,\infty)$ since it exists
bijective mappings $h: [0,\infty) \rightarrow (0,1]$ between these sets (e.g., $h(x)=e^{-x}$). However, a bounded set as $[0,1]$ is more neat to use.}
For multicell systems, $K_t(K_r-1)$ complex-valued parameters were used in \cite{Bjornson2010c}, which was improved to
$[0,1]$-parameters in \cite{Mochaourab2011a}.
Compared with this prior work, our new parametrization in Theorem \ref{theorem_beamforming_parametrization} generally requires much fewer parameters;
for instance, $K_t+K_r$ instead of $K_t(K_r-1)$ in MISO interference channels and other multicell systems with per-transmitter power constraints (i.e., $K_c=1, L=K_t$). In other words, the number of parameters increase linearly instead of quadratically. However, it is worth noting that the parametrizations in \cite{Zhang2010a,Shang2010a, Mochaourab2011a} are superior in the special case of $K_t=K_r=2$,
simply because Theorem \ref{theorem_beamforming_parametrization} handles any power constraint while only per-transmitter constraints are considered in \cite{Zhang2010a,Shang2010a, Mochaourab2011a}.

Main applications of Theorem \ref{theorem_beamforming_parametrization} are to search for parameter values iteratively or to select them heuristically.
An example of the former is the search algorithm in \cite{Zhang2010a}, which cannot guarantee global convergence but satisfy a necessary condition on optimality.
The well-known signal-to-leakage-and-noise ratio (SLNR) beamforming strategy in \cite{Zhang2008a,Tarighat2005a} corresponds to a certain heuristic selection of our parameters, and the regularized zero-forcing approach in \cite{Peel2005a} resembles the optimal structure in Theorem \ref{theorem_beamforming_parametrization}. Thus, the theorem explains why these strategies perform well and demonstrates that even better performance can be achieved by fine-tuning the parameters.

\section{Low-Complexity OFDMA Resource Allocation}
\label{section_distributed_precoding}

The resource allocation problem in \eqref{eq_P1} is generally NP-hard, making the optimal solution practically infeasible. We will therefore propose low-complexity strategies for OFDMA resource allocation that exploit the optimality properties in Section \ref{section_optimality_properties}. Despite the huge reduction in computational complexity, these strategies provide close-to-optimal performance in the measurement-based evaluation of Section \ref{section_numerical_examples}. In addition, the optimal multiplexing gain is achieved in certain scenarios.

Theorem \ref{theorem_beamforming_parametrization} parameterized the optimal solution to \eqref{eq_P1}, thus good performance can be achieved by judicious selection of the parameters $\lambda_{kc}$ and $\omega_l$. An important observation is that \eqref{eq_P1} allocates terminals over subcarrier as an implicit part of the optimization problem. In the parameterization, this scheduling is explicitly represented by having $\lambda_{kc}>0$ for all terminals $k$ that are scheduled on subcarrier $c$. Thus, heuristic parameter selection requires an efficient subcarrier scheduling mechanism. Herein, we adopt and extend the state-of-the-art ProSched algorithm from \cite{Fuchs2006a}.

\subsection{Centralized Resource Allocation}
\label{subsection_CVSINR}

For notational convenience, we only consider per-transmitter power constraints, weighted sum performance $f(
g_{1},\ldots,g_{K_r})=\sum_k \mu_k g_k$, and quality functions $g_{k}(\cdot)$ that can be decomposed as
\begin{equation} \label{eq_quality_function_decomposed}
g_{k} \left( \left\{ \textrm{SINR}^{\textrm{DL}}_{kc}
\right\}_{c=1}^{K_c} \right) = \sum_{c=1}^{K_c} \tilde{g}_{kc}
\left( \textrm{SINR}^{\textrm{DL}}_{kc} \right)
\end{equation}
where all $\tilde{g}_{kc}(\cdot)$ are concave functions. This
structure holds for both data rates, mean square errors (MSEs),
and symbol error rates (SERs), as will be shown below.

As the parametrization in Theorem \ref{theorem_beamforming_parametrization} builds upon virtual uplink optimization, we call our strategy \emph{centralized virtual SINR (CVSINR) resource allocation}. It is outlined as follows:

\begin{enumerate}
\item Consider weighted sum optimization $f( g_{1},\ldots,g_{K_r})=\sum_k \mu_k g_k$ for some collection of weights $\mu_k \geq 0$.
\item Allocate terminals $\widetilde{\mathcal{S}}_c \subseteq \mathcal{K}$ to subcarrier $c$ using an appropriate algorithm (e.g., ProSched \cite{Fuchs2006a}).
\item Set $\lambda_{kc}=\mu_k |\widetilde{\mathcal{S}}_c|/(\sigma_{kc}^2 \sum_{\bar{k} \in \widetilde{\mathcal{S}}_c} \mu_{\bar{k}}) $ if $\text{MS}_k$ is scheduled on subcarrier $c$, otherwise set $\lambda_{kc}=0$.
\item Set $\omega_l=K_c/q_l$ and calculate the signal correlation matrices $\vect{S}_{kc}$ using Theorem \ref{theorem_beamforming_parametrization}.
\item Rescale all $\vect{S}_{kc}$, according to Theorem \ref{theorem_power_usage}, to satisfy all power constraints (and some with equality).
\end{enumerate}

This CVSINR strategy allocates terminals to subcarriers and then calculates a single-stream beamforming strategy in compliance with the optimality properties in Section \ref{section_optimality_properties}. The parameters $\omega_l,\lambda_{kc}$ are selected heuristically, and the details are provided later in this section.

\subsection{Distributed Resource Allocation}
\label{subsection_DVSINR}

The drawback of any centralized strategy, including the proposed CVSINR strategy, is that resource allocation requires global CSI. In a system with many transmitters/receivers and many subcarriers, this requires huge amounts of backhaul signalling. In addition, joint CSI processing typically means large computational demands. Therefore, our main focus is to derive a low-complexity distributed version of CVSINR based on local CSI. It will be suboptimal in terms of performance, but have much more reasonable system requirements than centralized approaches.

Under single-stream beamforming, we have $\vect{S}_{kc} =
\vect{w}_{kc} \vect{w}_{kc}^H$ and divide the collective beamforming
vectors as $\vect{w}_{kc}=[\sqrt{p_{1kc}}\vect{v}_{1kc}^T \ldots \sqrt{p_{K_t kc}} \vect{v}_{K_t
kc}^T ]^T$. Here, $\vect{v}_{jkc} \in \mathbb{C}^{N_j \times 1}$ is the unit-norm beamforming vector and $p_{jkc} \geq 0$ is the power allocated by $\textrm{BS}_j$ for transmission to $\textrm{MS}_k$ (only non-zero for $k  \in \mathcal{D}_j$). With this notation, \eqref{eq_P1} becomes
\begin{align} \label{eq_P1_with_weighted_sum}
\maximize{\substack{
\{ \vect{v}_{jkc} \}_{j=1,k=1,c=1}^{K_t,K_r,K_c} \\
\{ p_{jkc} \}_{j=1,k=1,c=1}^{K_t,K_r,K_c}} } &\,\,  \,\,\,\,
\sum_{k=1}^{K_r} \mu_k \sum_{c=1}^{K_c} \tilde{g}_{kc} \left(
\textrm{SINR}^{\textrm{DL}}_{kc} \right) \\ \notag
\mathrm{subject}\,\,\mathrm{to} \,\,\,\,\,\,&\,\,  \,\,\,\,\,  p_{jkc}\geq 0, \|\vect{v}_{jkc}\|=1 \quad \forall j,k,c, \\ \notag
& \!\!\!\!\!\!\!\!\!\!\!\!\!\!\!\!\!\!\!\!\!\!\!\!\!\!\!\!\!\!\!\!\! \textrm{SINR}^{\textrm{DL}}_{kc} =
 \frac{ \big| \fracSum{j \in \mathcal{J}} \! \sqrt{p_{jkc}} \vect{h}_{jkc}^H \vect{D}_{jk} \vect{v}_{jkc} \big|^2}{\sigma_{kc}^2 \!+\!
 \fracSum{\bar{k} \in \mathcal{I}_k} \big| \fracSum{j  \in \mathcal{J}} \!\sqrt{p_{j\bar{k}c}} \vect{h}_{jkc}^H \vect{C}_{jk} \vect{D}_{j\bar{k}} \vect{v}_{j\bar{k}c} \big|^2} \,\,\, \forall k,c,\\ \notag
&\, \,\,\, \sum_{k=1}^{K_r} \sum_{c=1}^{K_c} p_{jkc}  \leq
q_j \,\,\,\, \forall j.
\end{align}

An important question is how to maximize $\textrm{SINR}^{\textrm{DL}}_{kc}$ in \eqref{eq_P1_with_weighted_sum} distributively using only local CSI.
Starting with the numerator, coherent signal reception can still be achieved, for instance by requiring that
$\sqrt{p_{jkc}} \vect{h}_{jkc}^H \vect{D}_{jk} \vect{v}_{jkc}$ should be positive real-valued for every transmitter.
Achieving coherent interference cancelation (i.e., that $|\sum_{j \in \mathcal{J}} \sqrt{p_{j\bar{k}c}} \vect{h}_{jkc}^H \vect{C}_{jk} \vect{D}_{j\bar{k}} \vect{v}_{j\bar{k}c}|^2$ is small without enforcing that every term is small) is more difficult under local CSI, if not impossible in large multiuser systems.\footnote{Iterative optimization can be used, but it requires backhaul signaling and is sensitive to CSI uncertainty, delays, etc.} Without coherent interference cancelation, there are few reasons for joint transmission; it is more power efficient and reliable to only use the transmitter with the strongest link, although somewhat more unbalanced interference patterns are generated. Therefore, each terminal is only served by one base station at each subcarrier in our distributed strategy---but different transmitters can be used on different subcarriers. This assumption greatly reduces the synchronization requirements, while the performance loss is small or even nonexistent (see Section \ref{section_numerical_examples}).

The resource allocation problem in \eqref{eq_P1_with_weighted_sum} can be divided into three parts: 1) Subcarrier allocation; 2) Power allocation $\{p_{jkc}\}$; and 3) Beamforming selection $\{\vect{v}_{jkc}\}$.
Our distributed strategy solves them sequentially, only requiring local CSI at each transmitter in each part (i.e.,
$\vect{h}_{jkc}$ is known at $\textrm{BS}_j$ for terminals $k\in \mathcal{C}_j$). In between each step, a small amount of signaling is used to tell each transmitter which terminals that are served by adjacent transmitter (to enable interference coordination). Next, we describe the three steps in detail.

\subsection{Step 1: Subcarrier Allocation}
\label{subsection_subcarrier_allocation}

The goal of this step is to select the scheduling set $\mathcal{S}_n(j,c)$ with terminals that are served by $\textrm{BS}_j$ at subcarrier $c$. The subscript $n$ denotes the current scheduling slot.
Observe that the performance of subcarriers is only coupled through the power constraints. Thus, it is reasonable to perform independent user scheduling on each subcarrier. The proposed scheme is a generalization of the ProSched algorithm in\cite{Fuchs2006a} and \cite{Fuchs2007a}, where the interference generated on terminals served by other transmitters,
\begin{equation}
\mathcal{A}_n(j,c) = \bigcup_{i \neq j} \left( \mathcal{S}_n(i,c) \cap \mathcal{C}_j \right),
\end{equation}
is also taken into consideration. For a given set $\mathcal{A}$, the scheduling metric for terminal $k$ is
\begin{equation}
\eta^{(\mathcal{S},\mathcal{A})}_{jkc}=\mu_k  \tilde{g}_{kc} \left( \frac{q_j \| \vect{h}_{jkc}^H \tilde{\vect{P}}^{(\mathcal{S},\mathcal{A})}_{jkc}\|^2  }{\sigma_{kc}^2 K_c |\mathcal{S}| }   \right)
\end{equation}
where $\tilde{\vect{P}}^{(\mathcal{S},\mathcal{A})}_{jkc}$ denotes the projection matrix onto the null space of channels for terminals in $(\mathcal{S} \cup \mathcal{A}) \setminus \{k\}$.
Thus, $\eta^{(\mathcal{S},\mathcal{A})}_{jkc}$ represents the performance with equal power allocation and zero-forcing precoding. To lower the computational complexity, the ProSched algorithm calculates $\tilde{\vect{P}}^{(\mathcal{S})}_{jkc}$ using an efficient approximation (see \cite{Fuchs2007a}). In the search for the scheduling set $\mathcal{S}$ with the highest sum metric
\begin{equation} \label{eq_scheduling_sum_metric}
 \bar{\eta}^{(\mathcal{S},\mathcal{A})}_{jc} = \sum_{k \in \mathcal{S}}  \eta^{(\mathcal{S},\mathcal{A})}_{jkc},
\end{equation}
 the ProSched algorithm avoids the complexity of evaluating it for every possible $\mathcal{S} \subseteq \mathcal{C}_j$ by performing a greedy tree search. Despite all simplifications, ProSched has shown good performance under reasonable complexity \cite{Fuchs2006a,Fuchs2007a}.
Our distributed ProSched algorithm exploits time correlation and selects $\mathcal{S}_n(j,c)$ as follows:

\begin{enumerate}
\item Start with $\mathcal{S}^{\textrm{tmp}}_n(j,c)=\mathcal{S}_{n-1}(j,c)$ and knowledge of $\mathcal{A}_{n-1}(j,c)$.

\item Use the "Tracking and Adaptivity"-procedure in \cite{Fuchs2007a} to add and remove terminals from $\mathcal{S}^{\textrm{tmp}}_n(j,c)$, while zero interference is generated to terminals in $\mathcal{A}_{n-1}(j,c)$. The final set needs to satisfy $|\mathcal{S}^{\textrm{tmp}}_n(j,c) \cup \mathcal{A}_{n-1}(j,c) |\leq N_j$.\footnote{A feature of the approximate zero-forcing precoding is that the sum metric will be non-zero also for $|\mathcal{S}^{\textrm{tmp}}_n(j,c) \cup \mathcal{A}_{n-1}(j,c) |> N_j$, but this corresponds to an interference-limited system and should be avoided.} The sum metric is evaluated using \eqref{eq_scheduling_sum_metric}.

\item Set $\mathcal{S}_n(j,c) = \mathcal{S}^{\textrm{tmp}}_n(j,c)$ and send it to central station.

\item Central station calculates $\mathcal{A}_n(j,c)$ and sends it to $\textrm{BS}_j$.
\end{enumerate}

The main difference from the original ProSched algorithm is the existence of $\mathcal{A}_{n-1}(j,c)$, which are terminals that $\textrm{BS}_j$ must coordinate interference towards. The algorithm exploits time correlation by taking new decisions based on previous ones; it tries to remove users from the selected set and then add other users. The user weights can updated between scheduling slots, although not reflected in our notation.
The initialization can be achieved in some arbitrary way, for example by selecting the strongest user as $\mathcal{S}_n(j,c) = \mathrm{argmax}_{k \in \mathcal{D}_j} \eta^{(\{k\},\emptyset)}_{jkc}$. The last step of the algorithm prepares for the next scheduling slot and is used in the next steps to adapt the precoding to the current subcarrier allocation.

\subsection{Step 2: Power Allocation}

The difficulty in distributed power allocation is that the interference powers generated by other transmitters are unknown. Fortunately, the proposed subcarrier allocation is designed to make
$|\mathcal{S}_{n}(j,c) \cup \mathcal{A}_{n}(j,c)|\leq N_j$ so that zero-forcing precoding exists.\footnote{Since the subcarrier allocation makes
$|\mathcal{S}^{\textrm{tmp}}_n(j,c) \cup \mathcal{A}_{n-1}(j,c)|\leq N_j$ with $\mathcal{A}_{n-1}(j,c)$ instead of $\mathcal{A}_{n}(j,c)$, it might occasionally happen that $|\mathcal{S}_{n}(j,c) \cup \mathcal{A}_{n}(j,c)|>N_j$. This is either handled by having a central mechanism that removes users or by ignoring the weakest unserved terminals in $\mathcal{A}_{n}(j,c)$ in the power allocation step. The latter will have limited impact on performance since most of the interference coordination comes from the beamforming directions and not from power allocation.}
Power allocation based on zero-forcing simplifications has been shown to provide accurate results (e.g., in the context of the ProSched algorithm), although better beamforming vectors are used in the end. Based on this assumption, the SINR of $\textrm{MS}_k$ at subcarrier $c$ becomes
\begin{equation} \label{eq_SINR_approx_power_alloc}
\textrm{SINR}^{\textrm{DL}}_{kc} = p_{jkc} \underbrace{\frac{ |\vect{h}_{jkc}^H
\vect{v}_{jkc}^{\textrm{ZF}} |^2}{\sigma^2_{kc}}}_{=\rho_{jkc}} \quad \forall k \in \mathcal{S}_n(j,c)
\end{equation}
where $\vect{v}_{jkc}^{\textrm{ZF}}$ is the unit-norm zero-forcing vector for terminals in $\mathcal{S}_{n}(j,c) \cup \mathcal{A}_{n}(j,c)$.
For fixed $\rho_{jkc}$, the distributed power allocation can be solved as follows.
\begin{lemma} \label{lemma_heuristic_power_allocation}
For a given transmitter index $j$, some given channel gain constants
$\rho_{jkc}>0$, and differentiable concave functions
$\widetilde{g}_{kc}(\cdot)$ with invertible derivatives, the power
allocation problem
\begin{equation}
\begin{split}
\maximize{p_{jkc}\geq0 \,\,\, \forall k \in \mathcal{S}_n(j,c), \, \forall c} &\quad
\sum_{c=1}^{K_c} \sum_{k \in \mathcal{S}_n(j,c)} \mu_k
\tilde{g}_{kc}(p_{jkc} \rho_{jkc}) \\
\mathrm{subject}\,\,\mathrm{to}\,\,\,\,\,\,\, & \quad \sum_{c=1}^{K_c} \sum_{k \in \mathcal{S}_n(j,c)}
 p_{jkc} \leq q_j
\end{split}
\end{equation}
is solved by
\begin{equation}
p_{jkc} = \max \left( \frac{1}{\rho_{jkc}}
\tilde{g}_{kc}'^{-1}\left( \frac{\nu}{\mu_k \rho_{jkc}} \right),0
\right)
\end{equation}
where $\frac{d}{dx} \widetilde{g}_{kc}(x) = \tilde{g}_{kc}'(x)$ and
$\nu \geq 0$ is selected to satisfy the constraint with equality.
\end{lemma}
\begin{IEEEproof}
This convex optimization problem is solved by standard Lagrangian
techniques \cite{Boyd2004a}.
\end{IEEEproof}

The distributed power allocation depends on the inverse of the
derivative of the terminal quality function $\widetilde{g}_{kc}(\cdot)$.
To exemplify the structure, we let the quality functions either describe the
data rate, MSE, or Chernoff bound\footnote{The
exact SER can also be used, but there are no closed-form expressions
for the inverse of its derivative.} on the SER for an uncoded $M$-ary
modulation:
\begin{align}
\tilde{g}^{\textrm{rate}}_{kc}(x) &= \log_2(1+x) &\Rightarrow
\,\, \tilde{g}_{kc}'^{-1}(y) &= \frac{1}{y}-1, \\
\tilde{g}^{\textrm{MSE}}_{kc}(x) &= -\frac{1}{1+x} &\Rightarrow
\,\, \tilde{g}_{kc}'^{-1}(y) &= \frac{1}{\sqrt{y}}-1, \\
\tilde{g}^{\textrm{cSER}}_{kc}(x) &= -\frac{M\!-\!1}{M} e^{-x z}
&\Rightarrow \,\, \tilde{g}_{kc}'^{-1}(y) &= \frac{1}{z} \log_e \!
\frac{(M\!-\!1)z}{My}. \label{eq_g_function_SER}
\end{align}
In \eqref{eq_g_function_SER}, $z=3/(M^2-1)$ for pulse amplitude
modulation (PAM), $z=\sin^2(\pi/M)$ for phase-shift keying (PSK),
and $z=3/(2M-2)$ for quadrature amplitude modulation (QAM).

For all the listed quality functions, the power
allocation in Lemma \ref{lemma_heuristic_power_allocation} has the
waterfilling behavior, which means that terminals receive more power on strong subcarriers than on weak.  In addition,
the system prioritizes terminals with large weights.
Some terminals might be allocated zero or negligible power (below some threshold $\tau$).
These terminals should immediately be removed from the scheduling set $\mathcal{S}_n(j,c)$, and adjacent base stations should be informed so that all $\mathcal{A}_n(j,c)$ can be adjusted.
This requires some extra signaling, but will avoid unnecessary interference coordination and improves the scheduling in the next slot.

\subsection{Step 3: Beamforming Selection}

The parametrization in Theorem \ref{theorem_beamforming_parametrization} provides the optimal structure of the beamforming directions.
Since at most one transmitter serves each terminal at each subcarrier, we have the following corollary.

\begin{corollary} \label{cor_beamforming_solution}
Assume that all $\mathcal{S}_{n}(j,c)$ are disjunct and that $\textrm{BS}_j$ has a per-transmitter power constraint of $q_j$.
For $\textrm{BS}_j$, the optimal beamforming direction to user $k \in \mathcal{S}_{n}(j,c)$ at subcarrier $c$ is given by
\begin{equation} \label{eq_parameterization_beamformer_distributed}
\vect{v}_{jkc} \!=\!
\frac{ \Big(  \omega_j \vect{I}_{N_j}  + \fracSum{\bar{k} \in \widetilde{\mathcal{I}}_k} \lambda_{\bar{k}c} \vect{h}_{j\bar{k}c} \vect{h}_{j\bar{k}c}^H
 \Big)^{\dagger} \vect{h}_{jkc} }{
 \Big\| \Big(  \omega_j \vect{I}_{N_j}  + \fracSum{\bar{k} \in \widetilde{\mathcal{I}}_k} \lambda_{\bar{k}c} \vect{h}_{j\bar{k}c} \vect{h}_{j\bar{k}c}^H
 \Big)^{\dagger} \vect{h}_{jkc} \Big\| }
\end{equation}
for some positive parameters $\omega_j$ and $\{\lambda_{\bar{k}c}\}_{\bar{k}=1,c=1}^{K_r,K_c}$.
\end{corollary}

\begin{table}
\vskip -1mm \caption{Distributed Virtual SINR (DVSINR) Resource Allocation} \vskip -5mm
\label{algorithm_distributed_solution}
\begin{lined}{8.2 cm} \vskip -1mm
\renewcommand{\labelenumi}{\theenumi:}
\begin{enumerate}

\item set power threshold $\tau$.

\item \textbf{for each} transmitter $j$ at scheduling slot $n$:

\item $\,\,\,$ set $\mathcal{S}^{\textrm{tmp}}_n(j,c)=\mathcal{S}_{n-1}(j,c)$.

\item $\,\,\,$ perform the "Tracking and Adaptivity"-procedure \cite{Fuchs2007a} \\ $\,\,\,$ on $\mathcal{S}^{\textrm{tmp}}_n(j,c)$ with the special rules in Section \ref{subsection_subcarrier_allocation}.

\item $\,\,\,$ set $\mathcal{S}_n(j,c) = \mathcal{S}^{\textrm{tmp}}_n(j,c)$ and send it to central station.

\item $\,\,\,$ attain $\mathcal{A}_n(j,c)$ from central station.

\item $\,\,\,$ calculate $p_{jkc}$ for $k \in \mathcal{S}_n(j,c)$ using Lemma \ref{lemma_heuristic_power_allocation}.

\item $\,\,\,$ remove terminals with $p_{jkc} < \tau$ from $\mathcal{S}_n(j,c)$.

\item $\,\,\,$ send updates of $\mathcal{S}_n(j,c)$ and attain updates of $\mathcal{A}_n(j,c)$.

\item $\,\,\,$ calculate $\omega_j$ and $\lambda_{kc}$ using \eqref{eq_heuristic_mu} and \eqref{eq_heuristic_lambda}.

\item $\,\,\,$ calculate $\vect{v}_{jkc}$ for $k \in \mathcal{S}_n(j,c)$ using Corollary \ref{cor_beamforming_solution}.

\item \textbf{end}
\end{enumerate}
\vskip 2mm
\end{lined}
\end{table}

To use Corollary \ref{cor_beamforming_solution}, the parameters $\omega_j$ and $\lambda_{\bar{k}c}$ need to be selected heuristically.
For this reason, recall that the parametrization is achieved using uplink-downlink duality. Thus, $\omega_j$ is inversely proportional to the SNR of the virtual uplink channels. As the parameter is user-independent, it is only affected by the transmit power of $\textrm{BS}_j$ and not of any noise parameter. We therefore select
\begin{equation} \label{eq_heuristic_lambda}
\omega^{(\textrm{heuristic})}_j = \frac{K_c}{q_j}.
\end{equation}
Next, we consider $\lambda_{\bar{k}c}$ and recall from \eqref{eq_impact_of_lambda} of Section \ref{section_optimality_properties} that $\lambda_{\bar{k}c}$ represents the priority of $\textrm{MS}_{\bar{k}}$, and whether or not the terminal is scheduled at subcarrier $c$. The best priority indicators that we have are the weights $\mu_{k}$ in \eqref{eq_P1_with_weighted_sum}, but we need to normalize them
based on which users are scheduled. Finally, $\lambda_{\bar{k}c}$ should be inversely proportional to the noise power $\sigma_{\bar{k}c}^2$ of $\textrm{MS}_{\bar{k}}$, since this term could not be included in the user-independent $\omega_j$-parameters. We therefore select
\begin{equation} \label{eq_heuristic_mu}
\lambda^{(\textrm{heuristic})}_{kc} =
\begin{cases}
\frac{ \mu_{k} }{ \sigma_{kc}^2 \sum_{\bar{k} \in \mathcal{S}\!\mathcal{A}_{n}(j,c)} \frac{ \mu_{\bar{k}} }{|\mathcal{S}\!\mathcal{A}_{n}(j,c)|} }, & \textrm{if } k \in \mathcal{S}\!\mathcal{A}_{n}(j,c)\\
0, & \textrm{otherwise}
\end{cases}
\end{equation}
where $\mathcal{S}\!\mathcal{A}_{n}(j,c)=\mathcal{S}_{n}(j,c) \cup \mathcal{A}_{n}(j,c)$, for notational convenience.
The normalization was performed such that the proposed scheme reduces to the well-studied SLNR beamforming strategy in \cite{Zhang2008a} if all user weights are identical and all noise terms are identical. Observe that different transmitters can have different heuristic values on $\lambda^{(\textrm{heuristic})}_{kc}$, representing the local terminal priority.

\subsection{Final Strategy}
The proposed distributed resource allocation strategy is summarized in Table \ref{algorithm_distributed_solution}. The strategy is named \emph{distributed virtual SINR (DVSINR) resource allocation}, since it based on optimization of virtual SINRs as in Lemma \ref{lemma_duality}. In addition, it reduces to the DVSINR beamforming scheme in \cite{Bjornson2010c} in certain single-carrier scenarios.

\begin{figure}[t!]
\begin{center}
\subfigure[Routes in Evaluation Scenario A.]
{\label{figure_random_locations}\includegraphics[width=\columnwidth]{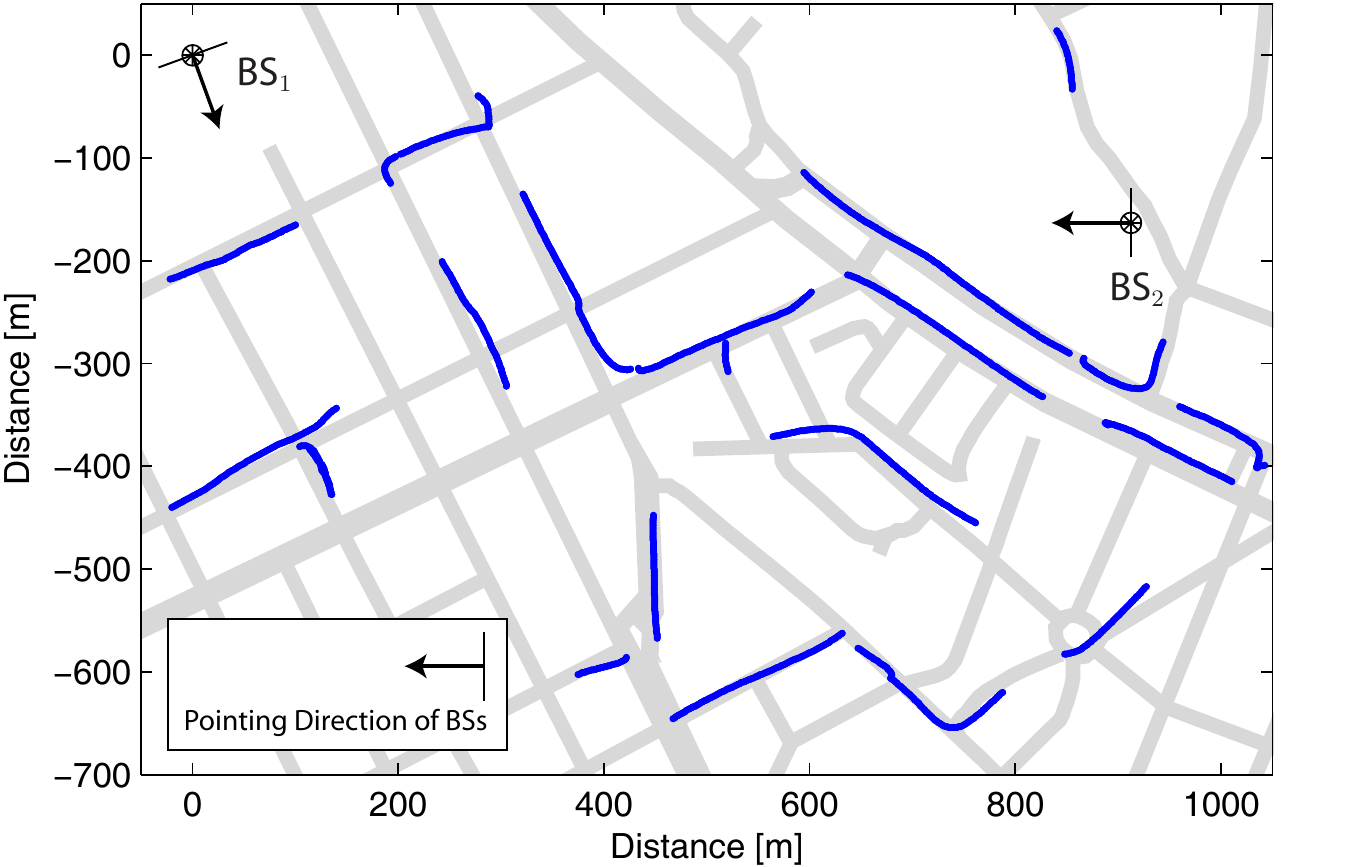}}\hfill
\subfigure[Routes in Evaluation Scenario B.]
{\label{figure_fixed_locations}\includegraphics[width=\columnwidth]{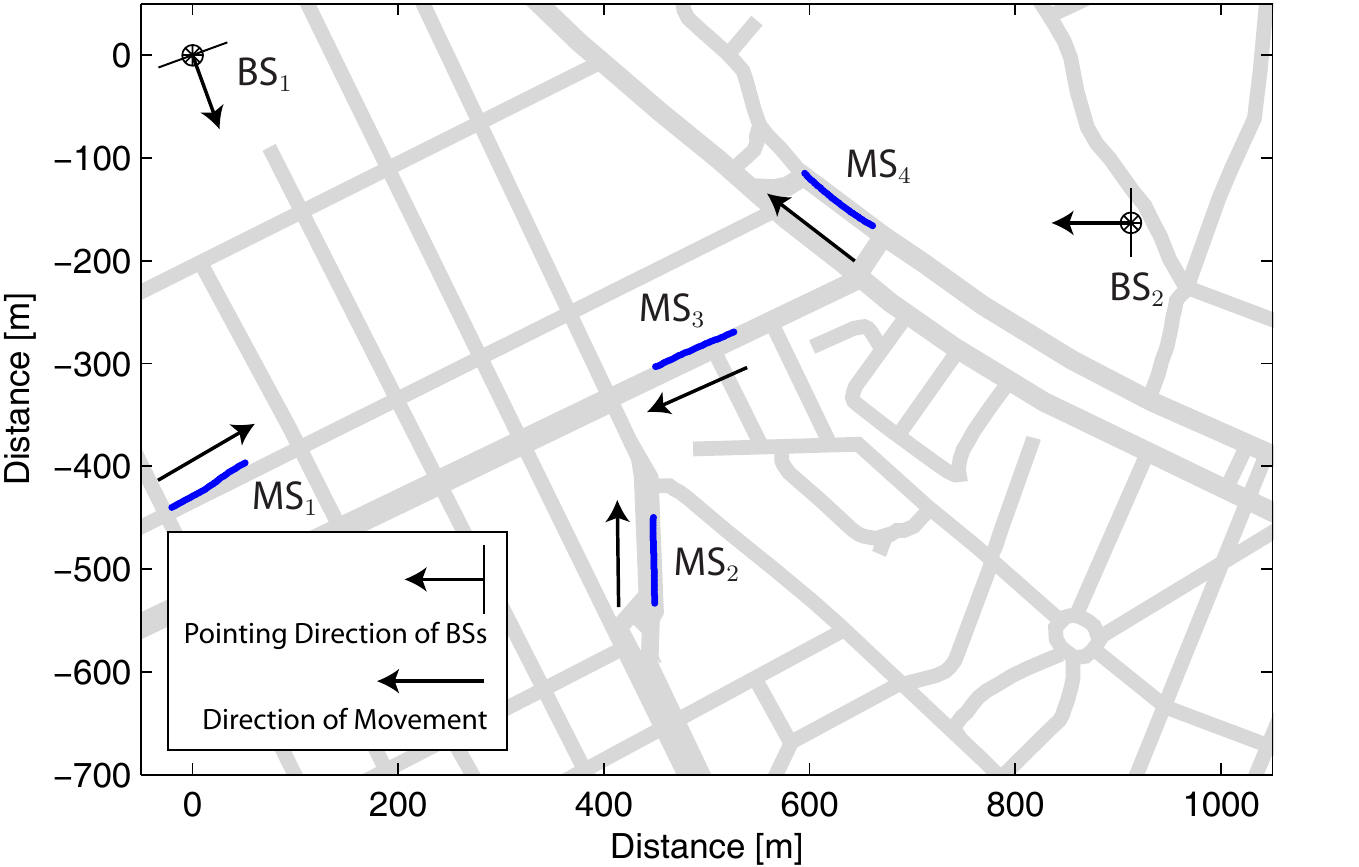}}
\caption{Downlink evaluation based on measurements in an urban
environment. Two four-antenna base stations are serving multiple
single-antenna terminals. These terminals are either randomly
located on the measured routes marked in Scenario A or move along
fixed routes as in Scenario B.}\label{figure_simulation_area}
\end{center}
\end{figure}

The proposed CVSINR and DVSINR strategies are both suboptimal, but for a given subcarrier scheduling they can achieve asymptotic optimality in terms
of multiplexing gain:

\begin{theorem} \label{theorem_multiplexing_gain}
Let $\mathcal{S}_c \subseteq \{1,\ldots,K_r\}$ be the terminals scheduled for transmission on subcarrier $c$.
If $|\mathcal{S}_c \cap \mathcal{C}_j|\leq N_j$ for all $j$ and $c$, then the CVSINR and DVSINR strategies
 achieve the multiplexing gain of $\sum_{c=1}^{K_c} |\mathcal{S}_c|$ (with probability one).
\end{theorem}
\begin{IEEEproof}
The theorem follows by the same approach as in \cite[Theorem
5]{Bjornson2010c} and exploits that random channels are linearly independent with probability one.
\end{IEEEproof}

This means that the weighted sum rate behaves as $(\sum_{c=1}^{K_c} |\mathcal{S}_c|) \log_2(P)+
\mathrm{constant}$ at high transmit power $P$. Thus, the absolute
performance losses (also called power offsets) of the CVSINR and DVSINR strategies are
bounded compared with the optimal solution, and the
relative loss goes to zero as $\mathrm{constant}/\log_2(P)$ with increasing transmit power.

\section{Measurement-Based Performance Evaluation}
\label{section_numerical_examples}

The theoretical performance of coordinated multicell transmission
has been thoroughly studied for single- and multi-carrier systems
(see e.g., \cite{Zhang2004a,Tolli2008a,Zheng2008a,Venturino2010a}).
Under Rayleigh fading channels and perfect synchronization, large
improvements over single-cell processing have been reported.
Especially, cell edge terminals benefit from inter-cell
interference coordination. However, results obtained from numerical
simulations are highly dependent on the assumptions in the
underlying channel models. For example, it is common to model the
channel characteristics between a terminal and multiple base
stations as uncorrelated, although correlation appears in practice
\cite{Jalden2007a}. Along with other idealized assumptions (e.g., on
fading distributions and path losses), such channel dependencies may
affect the performance of any multicell system.

The purpose of this section is to evaluate the performance of the low-complexity strategies in Section \ref{section_distributed_precoding} on realistic multicell scenarios based on channel measurements. We only consider a single subcarrier for computationally reasons (the optimal solution to \eqref{eq_P1} can be calculated systematically using \cite{Bjornson2012a}) and since our measurements are flat-fading. However, the subcarriers in weighted sum rate optimization are only coupled by the power constraints and OFDMA systems are known for giving almost flat power allocation over subcarriers \cite{Rhee2000a}. Thus, we expect our single-carrier results to be representative for general OFDMA systems.

The channel
data was collected in Stockholm, Sweden, using two base
stations\footnote{Channel data from a third base station, co-located
with $\textrm{BS}_1$, was also measured in \cite{Jalden2007a}. However, the overlap in coverage area between this base
station and $\textrm{BS}_1$ and $\textrm{BS}_2$ in Fig.~\ref{figure_simulation_area} is small, and
therefore it is not used herein.} with four-element uniform linear
arrays (ULAs) with $0.56\lambda$ antenna spacing and one terminal.
The terminal had a uniform circular array (UCA) with four
directional antennas, but herein we average the signal over its
antennas to create a single virtual omni-directional antenna. The
system bandwidth was 9.6 kHz at a carrier frequency in the 1800 MHz
band. The measurement environment can be characterized as typical
European urban with four to six story high stone buildings.
Fig.~\ref{figure_simulation_area} shows the measurement area with
roads illustrated in light gray and blue routes showing the GPS
coordinates of the terminal locations used for the simulations
herein. Further measurement details are available in
\cite{Jalden2007a}. The collected channel data is utilized to
generate two evaluation scenarios where terminals are moving
around in the area covered by both base stations:

\begin{itemize}
\item \textbf{Scenario A}: The performance behavior is evaluated over different random
terminal distributions. In each snapshot, terminals can be located
anywhere on the measured routes in Fig.~\ref{figure_random_locations} with uniform probability. To create balance,
two terminals are placed to have their strongest channel gains ($\|\vect{h}_{jk}\|^2$) from $\textrm{BS}_1$, and the same for $\textrm{BS}_2$.

\item \textbf{Scenario B}: To study the impact of coordination on
individual terminals, four terminals are now placed at certain locations and
moved as indicated in Fig.~\ref{figure_fixed_locations}.
\end{itemize}

\begin{figure}[t!]
\begin{center}
\subfigure[20 dBm output power per transmitter.]
{\label{figure_simulation_scheduling_20dBm}\includegraphics[width=\columnwidth]{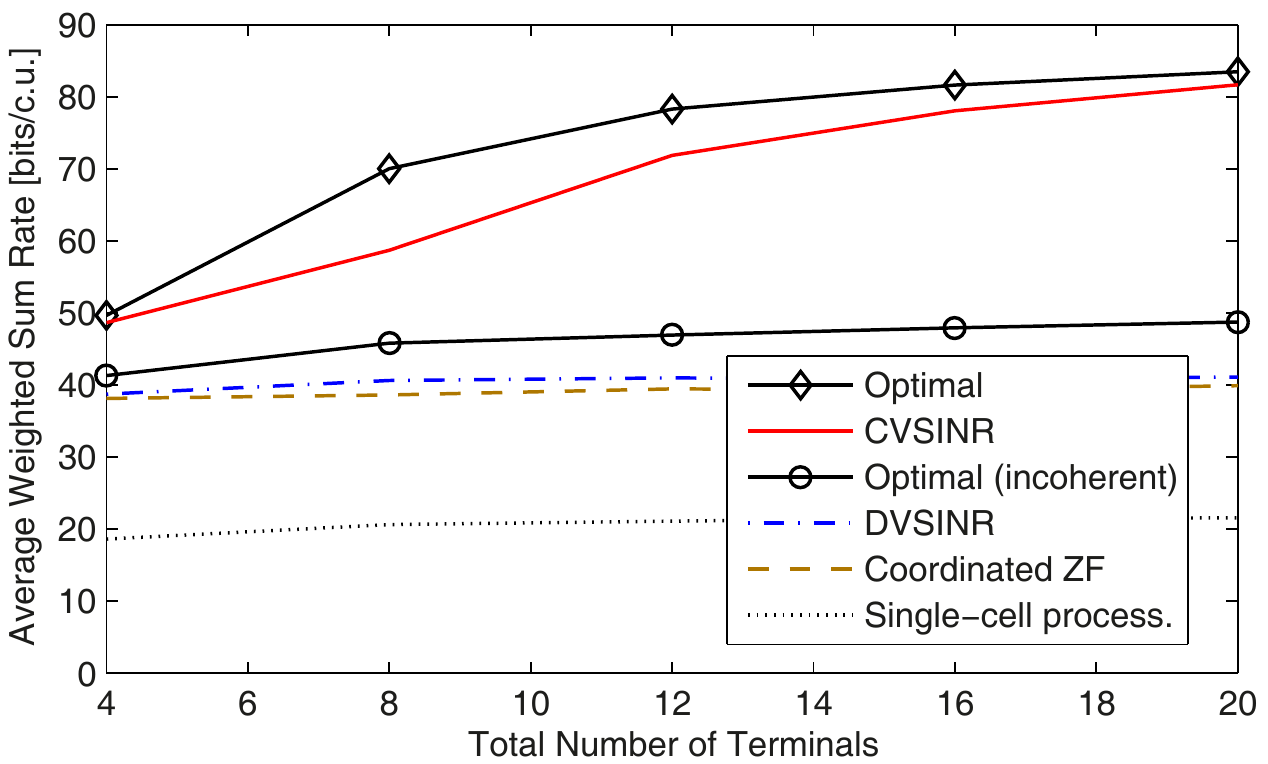}}\hfill
\subfigure[0 dBm output power per transmitter.]
{\label{figure_simulation_scheduling_0dBm}\includegraphics[width=\columnwidth]{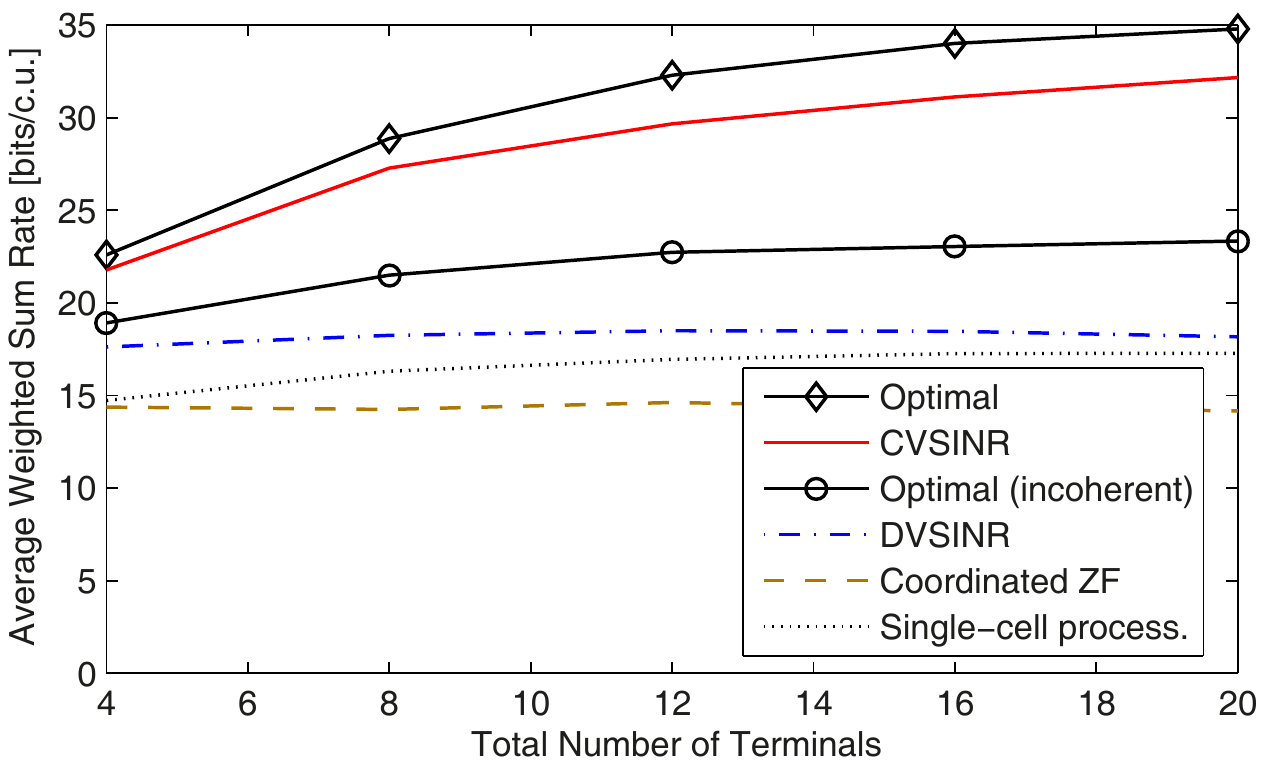}}
\caption{Average weighted sum rate (over random terminal locations) for Scenario A with different number of terminals.
The performance is shown for different resource allocation strategies.}\label{figure_simulation_scheduling}
\end{center}
\end{figure}

In both scenarios, the performance will be evaluated as a function
of the output power per base station (in dBm). The noise level is set to $-131$ dBm (i.e., thermal noise and a few dBs of transmitter noise)
 and the measured path losses from
the strongest base station varies between $-37$ dB and $-85$ dB for
different terminal locations in Fig.~\ref{figure_simulation_area}.

The performance measure is the weighted sum rate with
$\mu_k=c_w/\mathbb{E}\{ \log_2(1+\frac{K_t}{K_r \sigma_{kc}^2}
\max_j P_j \|\vect{h}_{jk}\|^2 ) \}$, where $c_w$ is a scaling
factor making $\sum_{k=1}^{K_r} \mu_k = K_r$. This represents
proportional fairness (with equal power allocation).
Six transmission strategies are compared:
\begin{enumerate}
\item The optimal solution (calculated using the framework in \cite{Bjornson2012a} and using the optimization software \texttt{CVX} \cite{cvx}.).

\item The optimal solution under incoherent interference reception, where base stations cannot cancel out each other's interference using joint transmission.
This case is relevant since very tight synchronization and long cyclic prefixes are necessary to enable coherent interference cancelation over wide areas \cite{Zhang2008a}. We
 represent it by replacing the SINR in \eqref{eq_P1_with_weighted_sum} with
\begin{equation}
\!\!\!\!\!\!\!\!\!\!\!\!\!\!\textrm{SINR}^{\textrm{DL-incoherent}}_{kc} =
 \frac{ \big| \fracSum{j \in \mathcal{J}} \sqrt{p_{jkc}} \vect{h}_{jkc}^H \vect{D}_{jk} \vect{v}_{jkc} \big|^2}{\sigma_{kc}^2 \!+\!
 \fracSum{\bar{k} \in \mathcal{I}_k} \fracSum{j \in \mathcal{J}} \big| \sqrt{p_{j\bar{k}c}} \vect{h}_{jkc}^H \vect{C}_{jk} \vect{D}_{j\bar{k}} \vect{v}_{j\bar{k}c} \big|^2}.
\end{equation}

\item Centralized virtual SINR (CVSINR) strategy, proposed in Section \ref{subsection_CVSINR}.

\item Distributed virtual SINR (DVSINR) strategy, proposed in Section \ref{subsection_DVSINR}.

\item Coordinated ZF precoding with single-cell scheduling using ProSched.

\item Single-cell processing as if there is only one cell in the system. The average out-of-cell interference is
included in the $\sigma_{kc}^2$-terms. The resource allocation is based on the DVSINR strategy (pretending that $K_t=1$).

\end{enumerate}

\begin{figure}[t!]
\includegraphics[width=\columnwidth]{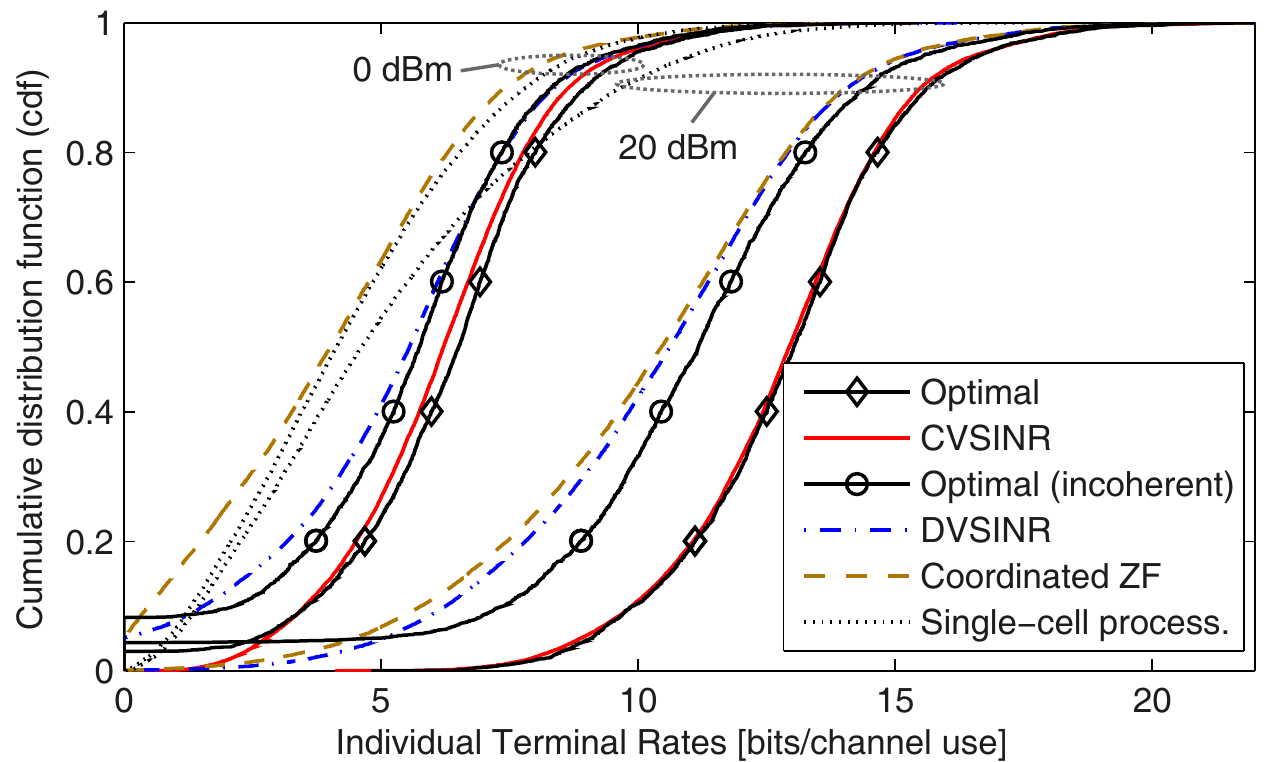}
\caption{Cumulative distribution function of the individual terminal
rates (over random terminal locations) for Scenario A with $K_r=4$. The
performance is shown for different resource allocation strategies at 0 dBm and
20 dBm output power.} \label{figure_simulation_cdf_random}
\end{figure}

\begin{figure}[t!]
\includegraphics[width=\columnwidth]{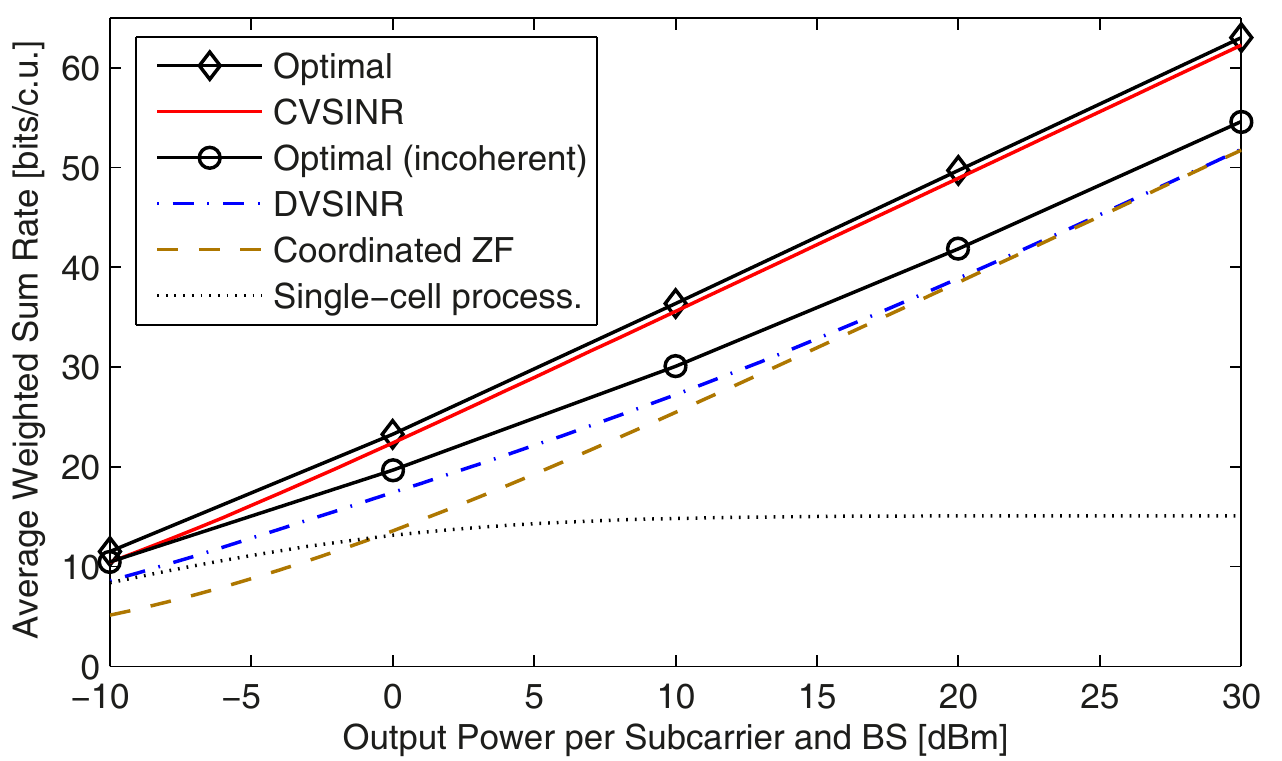}
\caption{Average weighted sum rate as a function of the output power for
Scenario B. The performance is shown for different resource allocation
strategies, including the proposed CVSINR and DVSINR strategies.}
\label{figure_simulation_weighted_fixed}
\end{figure}

\subsection{Results for Evaluation Scenario A}

The scheduling performance is evaluated in Fig.~\ref{figure_simulation_scheduling} over different random terminal locations, each used for 10 channel realizations.
The average weighted sum rate is given as a function of the total number of terminals at 20 dBm and 0 dBm output power per transmitter.
The proposed CVSINR strategy provides close-to-optimal performance, especially when the number of terminals increases.
The gap to the optimal solution is remarkably small, given that CVSINR is a simple combination of ProSched scheduling and heuristic use of the optimality properties derived in Section \ref{section_optimality_properties}---further parameter tweaking can certainly reduce the gap.
The distributed strategies (DVSINR and coordinated ZF) stabilize on about half the performance of the centralized schemes, representing that only half the number of terminals can be simultaneously accommodated. One might think that this is due to that only one transmitter serves each terminal, but the actual explanation is that (non-iterative) distributed schemes cannot achieve coherent interference cancelation. This is understood by the comparably small difference from Strategy 2 which includes joint transmission but have incoherent interference reception, and it confirms the discussion in Section \ref{subsection_DVSINR}.
All the studied centralized and distributed strategies provide improvements over single-cell processing, and the differences increase rapidly with the output power.

Next, we want to study how multicell coordination impacts the performance of each terminal and we set $K_r=4$ to make sure that all six strategies consider the same set of terminals. In Fig.~\ref{figure_simulation_cdf_random}, the cumulative distribution functions (cdfs) of
the individual terminal rates are given for output powers 0 dBm and 20 dBm.
The proposed CVSINR strategy is very close to the optimal solution, in particular at high output power.
The difference between the optimal solution
and the DVSINR strategy increases with the SNR, but the distributed
approach is close to the optimum under incoherent interference, which might be the most reasonable upper
bound in practice \cite{Zhang2008a}.
Both CVSINR and DVSINR provide great improvements over single-cell
processing---especially at high output power.
The coordinated ZF scheme performs poorly at low output power, but approaches
DVSINR at higher power.
To summarize, terminals that move
around in the cell clearly benefit from multicell coordination on the average
through better statistical properties. In Scenario B, we will
however see that terminals that are fixed at certain locations may
experience performance degradations.

\subsection{Results for Evaluation Scenario B}

For Scenario B, the average weighted sum rate (per channel use and
over 750 channel realizations) is shown in
Fig.~\ref{figure_simulation_weighted_fixed}.
Once again, the proposed CVSINR strategy provides close-to-optimal performance.
As in Scenario A, there is a clear gap to the distributed
approaches, explained by fewer degrees of freedom in the interference cancelation.
However, both DVSINR and coordinated ZF achieve the maximal multiplexing gain in certain scenarios (see Theorem
\ref{theorem_multiplexing_gain}), while the performance of
single-cell processing is bounded at high output power.
Observe that the major gain over single-cell processing comes from interference cancelation for terminals in $\mathcal{C}_j$;
the difference between DVSINR and optimal joint transmission is comparably small ($5-10$ dB).

\begin{figure}[t!]
\includegraphics[width=\columnwidth]{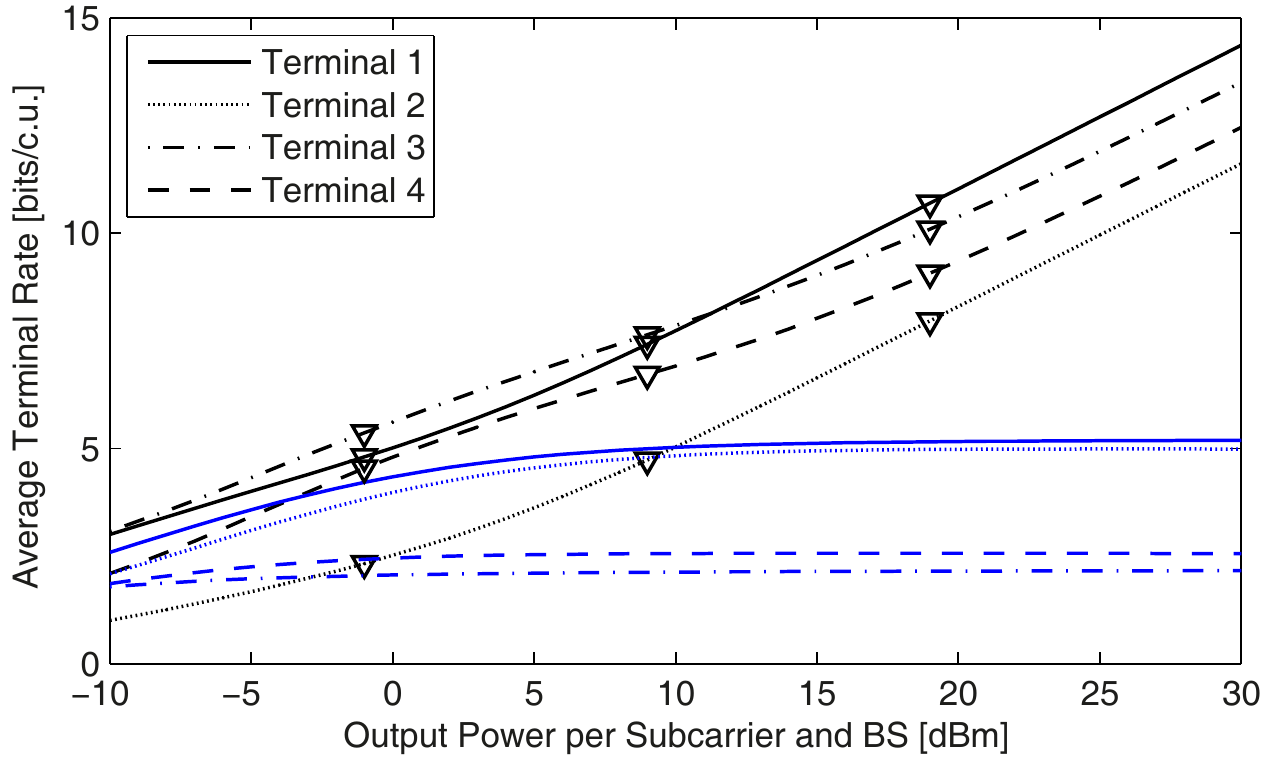}
\caption{Average individual terminal rates for Scenario B with and
without multicell coordination. The proposed DVSINR strategy
(triangles) is compared with single-cell processing.}
\label{figure_simulation_individual_fixed}
\end{figure}

\begin{figure}[t!]
\includegraphics[width=\columnwidth]{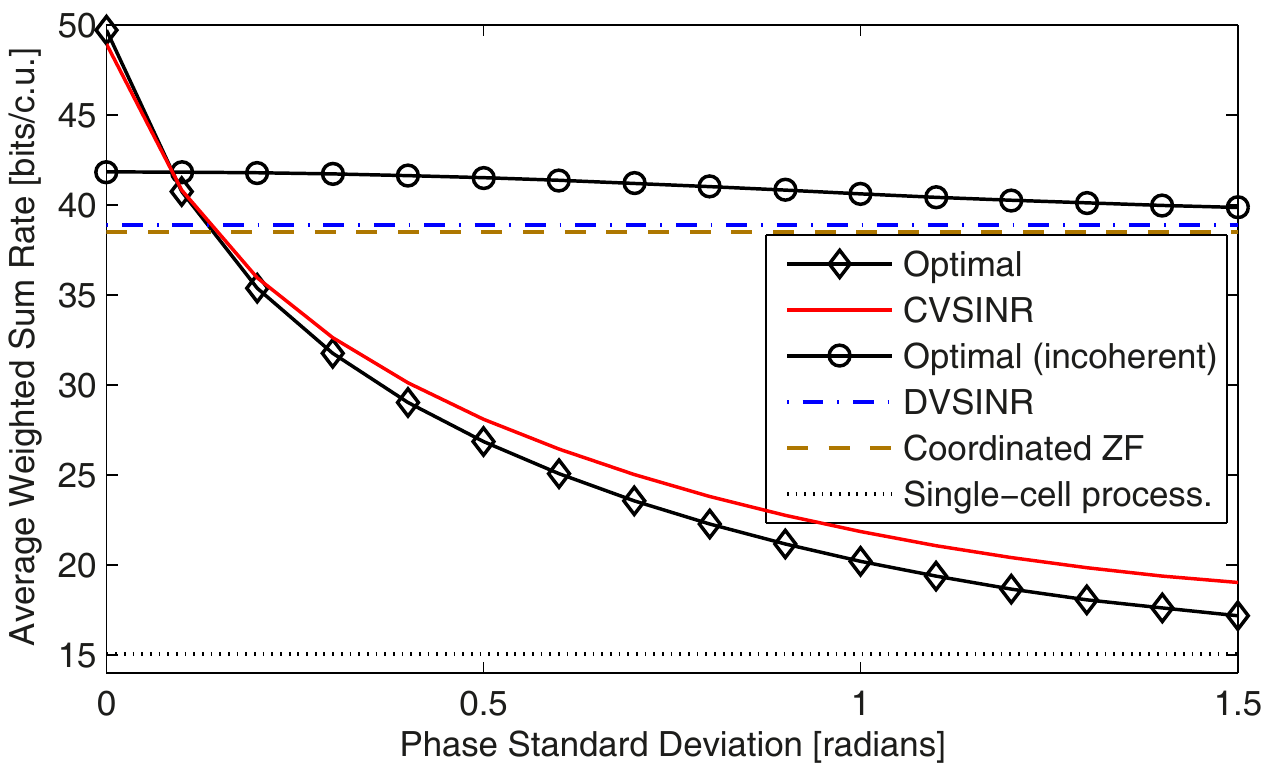}
\caption{Weighted sum rate for Scenario B as a function of phase
standard deviation $\sigma_{kc}$ (at 20 dBm output power). The
actual channels are modeled as
$\vect{h}_{jk}^{\textrm{actual}}=\vect{h}_{jk} e^{i \phi_{jk}}$,
where $\phi_{jk} \in \mathcal{N}(0,\sigma_{\phi}^2)$.}
\label{figure_simulation_sync_fixed}
\end{figure}

Fig.~\ref{figure_simulation_individual_fixed} shows the average
individual terminal rates for DVSINR (marked with
triangles) and single-cell processing. Interestingly, the increased
weighted sum rate with multicell coordination does not translate
into a monotonic improvement of all terminal rates. Terminal 3 has
almost equally strong channels from both base stations and therefore gain
substantially from interference coordination. However,
Terminal 2 has a very weak link to $\textrm{BS}_2$ and sees a
decrease in performance for output powers smaller than 10 dBm. This
is explained by $\textrm{BS}_1$ modifying its precoding to avoid interference at
Terminals 3 and 4. Thus, the common claim that multicell
coordination improves both the total throughput and the fairness is
not necessarily true in practice. However, Scenario A showed that as
terminals move around in the whole area, they will on average
benefit from multicell coordination.

The analysis has thus far considered perfect base station
synchronization, which cannot be guaranteed in practice due to CSI
uncertainty, hardware delays, clock drifts, insufficient cyclic prefixes, and minor channel
changes. We emulate these mismatches by letting the actual
channels be $\vect{h}_{jkc}^{\textrm{actual}}=\vect{h}_{jkc} e^{i
\phi_{jkc}}$ for some random phase deviations $\phi_{jkc} \in
\mathcal{N}(0,\sigma_{\phi}^2)$ (where $\sigma_{\phi}=0$ means perfect
synchronization). In Fig.~\ref{figure_simulation_sync_fixed}, the
average weighted sum rate is shown as a function of the phase
standard deviation $\sigma_{\phi}$ (at 20 dBm output power). The
optimal solution and the CVSINR strategy are very sensitive to synchronization errors as they rely on
coherent interference cancelation where interfering signals from different base stations
should cancel out perfectly. The DVSINR and coordinated ZF strategies are unaffected by such synchronization errors, and the gap to the
optimal schemes based on incoherent interference reception reduces with $\sigma_{\phi}$.
We conclude that very tight synchronization is required to gain from centralized multicell coordination with joint transmission.

\section{Conclusion}

A general multicell OFDMA resource allocation framework was introduced with
dynamic cooperation clusters that enables unified analysis of anything from interference channels to ideal network MIMO.
Joint precoding and scheduling optimization was considered using arbitrary monotonic utility functions and linear power constraints.
This problem is typically non-convex and NP-hard, but we proved three properties of the optimal solution:
1) Optimality of single-stream beamforming; 2) Conditions for full power usage; and 3) A precoding parametrization based on $K_r K_c +L$ real-valued parameters between zero and one. These properties greatly reduces the search space for optimal resource allocation.
To illustrate their usefulness, we proposed the centralized CVSINR strategy and the distributed DVSINR strategy. Both exploited the three optimality properties in conjunction with efficient ProSched subcarrier scheduling.

Contrary to previous work, the multicell performance was evaluated
on measured channels in a typical urban macro-cell scenario.
Substantial performance gains over single-cell processing were
observed for both CVSINR and DVSINR. The former is even close-to-optimal, while the latter performs closely to what can be expected from distributed schemes (since coherent interference cancelation is more or less impossible to achieve).
This is remarkable since both CVSINR and DVSINR are just simple applications of the derived optimality properties---further parameter tweaking and adaptation to special scenarios are possible. The performance evaluation also showed that joint transmission is very sensitive to synchronization errors and that multicell coordinated improves the average terminal performance, but that terminals in some parts of the cells can experience performance degradations.

\appendices

\section{Proof of Lemma \ref{lemma_duality}}
\label{appendix_proof_lemma_duality}

Using Theorem \ref{theorem_beamforming_optimality}, it is sufficient to consider solutions to \eqref{eq_P2} of the form $\vect{S}_{kc} = \vect{w}_{kc} \vect{w}_{kc}^H$.
Without loss of generality, we can select the phase of $\vect{w}_{kc}$ such that
$\vect{h}_{kc}^H \vect{D}_k \vect{w}_{kc}>0$. Under these conditions, the feasibility problem \eqref{eq_P2} can be expressed as
\begin{equation} \label{eq_P2_SOCP}
\begin{aligned}
\minimize{\{ \vect{w}_{kc} \}_{k=1,c=1}^{K_r,K_c} }\,\, & \,\, 0 \\
\mathrm{subject}\,\,\mathrm{to}\,\,\,\, & \,\,  \sum_{k=1}^{K_r} \sum_{c=1}^{K_c} \tr\{ \vect{Q}_l \vect{S}_{kc}\} \leq q_l \,\,\,\, \forall l, \\
 &  \!\!\!\!\!\!\!\!\!\!\!\!\!\!\!\!\!\!\!\!\!\!\!\!\!\!\!\!\!\! \left\| \begin{IEEEeqnarraybox*}[][c]{c}
\vect{h}_{kc}^H \vect{C}_{k} \vect{D}_{\mathcal{I}_k(1)} \vect{w}_{\mathcal{I}_k(1) c}\\
\vdots\\
\vect{h}_{kc}^H \vect{C}_{k} \vect{D}_{\mathcal{I}_k(|\mathcal{I}_k|)} \vect{w}_{\mathcal{I}_k(|\mathcal{I}_k|) c}\\
\sigma_{kc}%
\end{IEEEeqnarraybox*}\right\| \!
\leq\! \frac{\vect{h}_{kc}^H \vect{D}_k
\vect{w}_{kc}}{\sqrt{\gamma_{kc}}} \,\,\, \forall k,c.
\end{aligned}
\end{equation}
Since the QoS constraints are second order cones, and the power constraints are semi-definite, this problem is convex and
strong duality holds \cite{Boyd2004a}. To derive the Lagrange dual problem, we
rewrite the Lagrangian of \eqref{eq_P2_SOCP} similarly to
\cite[Proposition 1]{Yu2007a}:
\begin{equation} \label{eq_lagrangian_P2}
\begin{split}
&\mathcal{L} \left( \{ \vect{w}_{kc} \}_{k=1,c=1}^{K_r,K_c},\boldsymbol{\omega},\{\boldsymbol{\lambda}_c\}_{c=1}^{K_c} \right) =
\Big( \sum_{k=1}^{K_r} \sum_{c=1}^{K_c} \lambda_{kc}
\sigma_{kc}^2  -
\sum_{l=1}^{L} \omega_l q_l \Big) \\
&+\sum_{k=1}^{K_r} \sum_{c=1}^{K_c} \vect{w}_{kc}^H
\Big( \sum_{l=1}^{L}  \omega_l \vect{Q}_l + \fracSum{\bar{k} \in \widetilde{\mathcal{I}}_k} \lambda_{\bar{k}c}
\vect{D}_k^H \vect{C}_{\bar{k}}^H \vect{h}_{\bar{k}c} \vect{h}_{\bar{k}c}^H \vect{C}_{\bar{k}} \vect{D}_k \\ & \quad\quad\quad\quad\quad\quad\quad\quad\quad-
\frac{\lambda_{kc}}{\gamma_{kc}} \vect{D}_k^H \vect{h}_{kc} \vect{h}_{kc}^H \vect{D}_k
\Big) \vect{w}_{kc}.
\end{split}
\end{equation}
Minimizing with respect to $\{ \vect{w}_{kc} \}_{k=1,c=1}^{K_r,K_c}$ gives a finite solution only if
\begin{equation} \label{eq_dual_constraint}
\begin{split}
\Big( \sum_{l=1}^L \omega_l \vect{Q}_l + \fracSum{\bar{k} \in
\widetilde{\mathcal{I}}_k} &\lambda_{\bar{k}c}
 \vect{D}_k^H  \vect{C}_{\bar{k}}^H \vect{h}_{\bar{k}c} \vect{h}_{\bar{k}c}^H \vect{C}_{\bar{k}} \vect{D}_k \\ &-
\frac{\lambda_{kc}}{\gamma_{kc}}  \vect{D}_k^H \vect{h}_{kc} \vect{h}_{kc}^H \vect{D}_k
\Big) \succeq \vect{0} \quad \forall k,c,
\end{split}
\end{equation}
and the minimum is achieved by  $\vect{w}_{kc}=\vect{0}$. Using \cite[Lemma 1]{Yu2007a}, this dual feasibility constraint can be rewritten as
\begin{equation} \label{eq_rewrite_with_receive_beamformer}
\begin{split}
\gamma_{kc} &\geq \lambda_{kc} \vect{h}_{kc}^H \vect{D}_k
\Big( \sum_{l}  \omega_l\vect{Q}_l + \fracSum{\bar{k} \in
\widetilde{\mathcal{I}}_k} \lambda_{\bar{k}c}
 \vect{D}_k^H \vect{C}_{\bar{k}}^H \vect{h}_{\bar{k}c} \vect{h}_{\bar{k}c}^H \vect{C}_{\bar{k}} \vect{D}_k \\
 &\quad\quad\quad\quad\quad\quad\quad\quad\quad-\frac{\lambda_{kc}}{\gamma_{kc}}  \vect{D}_k^H \vect{h}_{kc} \vect{h}_{kc}^H \vect{D}_k
\Big)^{\dagger} \vect{D}_k^H \vect{h}_{kc} \\
&= \max_{\bar{\vect{w}}_{kc}} \frac{ \lambda_{kc} \bar{\vect{w}}_{kc}^H
\vect{D}_k^H \vect{h}_{kc} \vect{h}_{kc}^H \vect{D}_k
\bar{\vect{w}}_{kc}}{ \bar{\vect{w}}_{kc}^H  (
\fracSum{l}  \omega_l \vect{Q}_l \!+\!\! \fracSum{\bar{k} \in
\widetilde{\mathcal{I}}_k} \lambda_{\bar{k}c} \vect{D}_k^H \vect{C}_{\bar{k}}^H
\vect{h}_{\bar{k}c} \vect{h}_{\bar{k}c}^H \vect{C}_{\bar{k}} \vect{D}_k )
 \bar{\vect{w}}_{kc}}
\end{split}
\end{equation}
where the equality follows from introducing maximization\footnote{In general, $( \sum_l \vect{Q}_l
\!+\!\! \sum_{\bar{k} \in \widetilde{\mathcal{I}}_k} \lambda_{\bar{k}c} \vect{D}_k^H \vect{C}_{\bar{k}}^H
\vect{h}_{\bar{k}c} \vect{h}_{\bar{k}c}^H \vect{C}_{\bar{k}} \vect{D}_k)$ is rank-deficient, thus it might seem like
there are solutions to \eqref{eq_rewrite_with_receive_beamformer}
that achieve infinite utility. However, this is not the case as such
solutions require $\vect{h}_{kc}^H \vect{D}_k \bar{\vect{w}}_{kc}
\neq 0$ and from \eqref{eq_dual_constraint} it is clear that the
denominator is non-zero for such solutions.} over an extra variable $\bar{\vect{w}}_{kc} \in \mathbb{C}^{N \times 1}$. Observe
that this maximization corresponds to a linear
MMSE receiver with optimum at
\begin{equation} \label{eq_linear_MMSE_receiver}
\bar{\vect{w}}_{kc}^* = \lambda_{kc} \Big( \sum_{l=1}^{L}  \omega_l \vect{Q}_l  \!+\!
\fracSum{\bar{k} \in \widetilde{\mathcal{I}}_k} \lambda_{\bar{k}c}
\vect{D}_k^H \vect{C}_{\bar{k}}^H \vect{h}_{\bar{k}c}
\vect{h}_{\bar{k}c}^H \vect{C}_{\bar{k}} \vect{D}_k \Big)^{\dagger}
\vect{D}_k^H \vect{h}_{kc}.
\end{equation}
By plugging $\vect{w}_{kc}=\vect{0}$ into
\eqref{eq_lagrangian_P2}, we achieve a Lagrange dual problem to
\eqref{eq_P2_SOCP} and \eqref{eq_P2}:
\begin{align} \label{eq_dual_to_P2_alternative}
\maximize{\boldsymbol{\omega}, \{ \boldsymbol{\lambda}_c\}_{c=1}^{K_c}}\,\, & \,\,
\sum_{k=1}^{K_r} \sum_{c=1}^{K_c} \lambda_{kc}
\sigma_{kc}^2  -
\sum_{l=1}^{L} \omega_l q_l \\ \notag
\mathrm{subject}\,\,\mathrm{to}\,\, & \,\, \lambda_{kc} \geq
0, \, \omega_l \geq 0 \,\,\,  \forall k,c,l, \\ \notag
& \!\!\!\!\!\!\!\!\!\!\!\!\max_{\bar{\vect{w}}_{kc}} \,
\textrm{SINR}^{\textrm{VUL}}_{kc}(\bar{\vect{w}}_{1
c},\ldots,\bar{\vect{w}}_{K_r c}, \boldsymbol{\omega}, \boldsymbol{\lambda}_c)
\leq \gamma_{kc}, \,\,\, \forall k,c.
\end{align}
At the optimum, all SINR constraints are active
(otherwise we could increase some $\lambda_{kc}$ and thereby increase the
cost function), thus we replace them with equality constraints and achieve \eqref{eq_dual_to_P2}.

Finally, to find a relationship between optimal
$\bar{\vect{w}}_{kc}$ and $\vect{w}_{kc}$ we consider the stationary
principle of the Lagrangian in \eqref{eq_lagrangian_P2}:
\begin{equation}
\begin{split}
\vect{0} &= \frac{\partial \mathcal{L}}{\partial \vect{w}_{kc}} \\
&= 2 \Big( \sum_{l=1}^{L}  \omega_l \vect{Q}_l + \fracSum{\bar{k} \in \widetilde{\mathcal{I}}_k} \lambda_{\bar{k}c}
\vect{D}_k^H \vect{C}_{\bar{k}}^H \vect{h}_{\bar{k}c} \vect{h}_{\bar{k}c}^H \vect{C}_{\bar{k}} \vect{D}_k \\ & \quad\quad\quad\quad\quad\quad\quad -
\frac{\lambda_{kc}}{\gamma_{kc}} \vect{D}_k^H \vect{h}_{kc} \vect{h}_{kc}^H \vect{D}_k
\Big) \vect{w}_{kc}.
\end{split}
\end{equation}
By defining the scalar $d_{kc} = \lambda_{kc} \vect{h}_{kc}^H \vect{D}_k
\vect{w}_{kc} /\gamma_{kc}$, we achieve
\begin{equation} \label{eq_dual_beamforming}
\vect{w}_{kc} = d_{kc} \Big( \sum_{l=1}^{L}  \omega_l \vect{Q}_l  \!+\!
\fracSum{\bar{k} \in \widetilde{\mathcal{I}}_k} \lambda_{\bar{k}c}
\vect{D}_k^H \vect{C}_{\bar{k}}^H \vect{h}_{\bar{k}c}
\vect{h}_{\bar{k}c}^H \vect{C}_{\bar{k}} \vect{D}_k \Big)^{\!\dagger}
\vect{D}_k^H \vect{h}_{kc}
\end{equation}
and identify $\bar{\vect{w}}_{kc}$ from
\eqref{eq_linear_MMSE_receiver} (which is unique up to scaling).

\section*{Acknowledgement}

The authors would like to thank Dr.~Per Zetterberg for providing the channel measurements and for his valuable guidance on their details.

\bibliographystyle{IEEEtran}
\bibliography{IEEEabrv,refs}

\begin{thebibliography}{10}
\providecommand{\url}[1]{#1}
\csname url@samestyle\endcsname
\providecommand{\newblock}{\relax}
\providecommand{\bibinfo}[2]{#2}
\providecommand{\BIBentrySTDinterwordspacing}{\spaceskip=0pt\relax}
\providecommand{\BIBentryALTinterwordstretchfactor}{4}
\providecommand{\BIBentryALTinterwordspacing}{\spaceskip=\fontdimen2\font plus
\BIBentryALTinterwordstretchfactor\fontdimen3\font minus
  \fontdimen4\font\relax}
\providecommand{\BIBforeignlanguage}[2]{{%
\expandafter\ifx\csname l@#1\endcsname\relax
\typeout{** WARNING: IEEEtran.bst: No hyphenation pattern has been}%
\typeout{** loaded for the language `#1'. Using the pattern for}%
\typeout{** the default language instead.}%
\else
\language=\csname l@#1\endcsname
\fi
#2}}
\providecommand{\BIBdecl}{\relax}
\BIBdecl

\bibitem{Jiang2007a}
M.~Jiang and L.~Hanzo, ``Multiuser {MIMO-OFDM} for next-generation wireless
  systems,'' \emph{Proc. {IEEE}}, vol.~95, no.~7, pp. 1430--1469, 2007.

\bibitem{Gesbert2007a}
D.~Gesbert, M.~Kountouris, R.~Heath, C.-B. Chae, and T.~S\"alzer, ``Shifting
  the {MIMO} paradigm,'' \emph{{IEEE} Signal Process. Mag.}, vol.~24, no.~5,
  pp. 36--46, 2007.

\bibitem{Venkatesan2007a}
S.~Venkatesan, A.~Lozano, and R.~Valenzuela, ``Network {MIMO}: Overcoming
  intercell interference in indoor wireless systems,'' in \emph{Proc.~IEEE
  ACSSC'07}, 2007, pp. 83--87.

\bibitem{Parkvall2008a}
S.~Parkvall, E.~Dahlman, A.~Furusk\"ar, Y.~Jading, M.~Olsson, S.~W\"anstedt,
  and K.~Zangi, ``{LTE}-advanced - evolving {LTE} towards {IMT}-advanced,'' in
  \emph{Proc.~IEEE VTC'08-Fall}, 2008.

\bibitem{Shamai2001a}
S.~Shamai and B.~Zaidel, ``Enhancing the cellular downlink capacity via
  co-processing at the transmitting end,'' in \emph{Proc.~IEEE VTC'01-Spring},
  vol.~3, 2001, pp. 1745--1749.

\bibitem{Zhang2004a}
H.~Zhang and H.~Dai, ``Cochannel interference mitigation and cooperative
  processing in downlink multicell multiuser {MIMO} networks,'' \emph{{EURASIP}
  J. Wirel. Commun. Netw.}, vol.~2, pp. 222--235, 2004.

\bibitem{Karakayali2006a}
M.~Karakayali, G.~Foschini, and R.~Valenzuela, ``Network coordination for
  spectrally efficient communications in cellular systems,'' \emph{{IEEE}
  Wireless Commun. Mag.}, vol.~13, no.~4, pp. 56--61, 2006.

\bibitem{Weingarten2006a}
H.~Weingarten, Y.~Steinberg, and S.~Shamai, ``The capacity region of the
  {Gaussian} multiple-input multiple-output broadcast channel,'' \emph{{IEEE}
  Trans. Inf. Theory}, vol.~52, no.~9, pp. 3936--3964, 2006.

\bibitem{Marsch2008a}
P.~Marsch and G.~Fettweis, ``On multicell cooperative transmission in
  backhaul-constrained cellular systems,'' \emph{Ann. Telecommun.}, vol.~63,
  pp. 253--269, 2008.

\bibitem{Simeone2009a}
O.~Simeone, O.~Somekh, H.~V. Poor, and S.~Shamai, ``Downlink multicell
  processing with limited-backhaul capacity,'' \emph{EURASIP J. on Adv. in
  Signal Process.}, 2009.

\bibitem{Bjornson2009e}
E.~Bj{\"{o}}rnson and B.~Ottersten, ``On the principles of multicell precoding
  with centralized and distributed cooperation,'' in \emph{Proc.~WCSP'09},
  2009.

\bibitem{Tolli2008a}
A.~T\"{o}lli, M.~Codreanu, and M.~Juntti, ``Cooperative {MIMO-OFDM} cellular
  system with soft handover between distributed base station antennas,''
  \emph{{IEEE} Trans. Wireless Commun.}, vol.~7, no.~4, pp. 1428--1440, 2008.

\bibitem{Boche2002a}
H.~Boche and M.~Schubert, ``A general duality theory for uplink and downlink
  beamforming,'' in \emph{Proc.~IEEE VTC'02-Fall}, 2002, pp. 87--91.

\bibitem{Wiesel2006a}
A.~Wiesel, Y.~Eldar, and S.~Shamai, ``Linear precoding via conic optimization
  for fixed {MIMO} receivers,'' \emph{{IEEE} Trans. Signal Process.}, vol.~54,
  no.~1, pp. 161--176, 2006.

\bibitem{Yu2007a}
W.~Yu and T.~Lan, ``Transmitter optimization for the multi-antenna downlink
  with per-antenna power constraints,'' \emph{{IEEE} Trans. Signal Process.},
  vol.~55, no.~6, pp. 2646--2660, 2007.

\bibitem{Dahrouj2010a}
H.~Dahrouj and W.~Yu, ``Coordinated beamforming for the multicell multi-antenna
  wireless system,'' \emph{{IEEE} Trans. Wireless Commun.}, vol.~9, no.~5, pp.
  1748--1759, 2010.

\bibitem{Liu2011a}
Y.-F. Liu, Y.-H. Dai, and Z.-Q. Luo, ``Coordinated beamforming for {MISO}
  interference channel: Complexity analysis and efficient algorithms,''
  \emph{{IEEE} Trans. Signal Process.}, vol.~59, no.~3, pp. 1142--1157, 2011.

\bibitem{Brehmer2010a}
J.~Brehmer and W.~Utschick, ``Optimal interference management in multi-antenna,
  multi-cell systems,'' in \emph{Proc.~Int. Zurich Seminar on Commun.}, 2010,
  pp. 134--137.

\bibitem{Bjornson2012a}
E.~Bj{\"{o}}rnson, G.~Zheng, M.~Bengtsson, and B.~Ottersten, ``Robust monotonic
  optimization framework for multicell {MISO} systems,'' \emph{{IEEE} Trans.
  Signal Process.}, 2011, submitted, arXiv:1104.5240v2.

\bibitem{Tolli2009b}
A.~T\"{o}lli, H.~Pennanen, and P.~Komulainen, ``On the value of coherent and
  coordinated multi-cell transmission,'' in \emph{Proc.~IEEE ICC'09}, 2009.

\bibitem{Zheng2008a}
G.~Zheng, K.-K. Wong, and T.-S. Ng, ``Throughput maximization in linear
  multiuser {MIMO-OFDM} downlink systems,'' \emph{{IEEE} Trans. Veh. Commun.},
  vol.~57, no.~3, pp. 1993--1998, 2008.

\bibitem{Venturino2010a}
L.~Venturino, N.~Prasad, and X.~Wang, ``Coordinated linear beamforming in
  downlink multi-cell wireless networks,'' \emph{{IEEE} Trans. Wireless
  Commun.}, vol.~9, no.~4, pp. 1451--1461, 2010.

\bibitem{Jalden2007a}
N.~Jald{\'e}n, P.~Zetterberg, B.~Ottersten, and L.~Garcia, ``Inter- and
  intrasite correlations of large-scale parameters from macrocellular
  measurements at {1800 MHz},'' \emph{{EURASIP} J. Wirel. Commun. Netw.}, 2007.

\bibitem{Jorswieck2008b}
E.~Jorswieck, E.~Larsson, and D.~Danev, ``Complete characterization of the
  {Pareto} boundary for the {MISO} interference channel,'' \emph{{IEEE} Trans.
  Signal Process.}, vol.~56, no.~10, pp. 5292--5296, 2008.

\bibitem{Bjornson2010c}
E.~Bj{\"{o}}rnson, R.~Zakhour, D.~Gesbert, and B.~Ottersten, ``Cooperative
  multicell precoding: Rate region characterization and distributed strategies
  with instantaneous and statistical {CSI},'' \emph{{IEEE} Trans. Signal
  Process.}, vol.~58, no.~8, pp. 4298--4310, 2010.

\bibitem{Shang2010a}
X.~Shang, B.~Chen, and H.~V. Poor, ``Multiuser {MISO} interference channels
  with single-user detection: Optimality of beamforming and the achievable rate
  region,'' \emph{{IEEE} Trans. Inf. Theory}, vol.~57, no.~7, pp. 4255--4273,
  2011.

\bibitem{Zhang2010a}
R.~Zhang and S.~Cui, ``Cooperative interference management with {MISO}
  beamforming,'' \emph{{IEEE} Trans. Signal Process.}, vol.~58, no.~10, pp.
  5450--5458, 2010.

\bibitem{Mochaourab2011a}
R.~Mochaourab and E.~Jorswieck, ``Optimal beamforming in interference networks
  with perfect local channel information,'' \emph{{IEEE} Trans. Signal
  Process.}, vol.~59, no.~3, pp. 1128--1141, 2011.

\bibitem{Bjornson2010d}
E.~Bj{\"{o}}rnson, M.~Bengtsson, and B.~Ottersten, ``Optimality properties and
  low-complexity solutions to coordinated multicell transmission,'' in
  \emph{Proc.~IEEE GLOBECOM'10}, 2010.

\bibitem{Zhang2004b}
Y.~Zhang and K.~Letaief, ``Multiuser adaptive subcarrier-and-bit allocation
  with adaptive cell selection for {OFDM} systems,'' \emph{{IEEE} Trans.
  Wireless Commun.}, vol.~3, no.~5, pp. 1566--1575, 2004.

\bibitem{Fuchs2006a}
M.~Fuchs, G.~D. Galdo, and M.~Haardt, ``Low complexity spatial scheduling
  {ProSched} for {MIMO} systems with multiple base stations and a central
  controller,'' in \emph{Proc.~ITG Workshop on Smart Antennas}, 2006.

\bibitem{Xu2010a}
X.~Xu, X.~Chen, and J.~Li, ``Handover mechanism in coordinated multi-point
  transmission/reception system,'' \emph{ZTE Communications}, no.~1, pp.
  31--35, 2010.

\bibitem{Goldsmith2003a}
A.~Goldsmith, S.~Jafar, N.~Jindal, and S.~Vishwanath, ``Capacity limits of
  {MIMO} channels,'' \emph{{IEEE} J. Sel. Areas Commun.}, vol.~21, no.~5, pp.
  684--702, 2003.

\bibitem{Zhang2008a}
H.~Zhang, N.~Mehta, A.~Molisch, J.~Zhang, and H.~Dai, ``Asynchronous
  interference mitigation in cooperative base station systems,'' \emph{{IEEE}
  Trans. Wireless Commun.}, vol.~7, no.~1, pp. 155--165, 2008.

\bibitem{Boyd2004a}
S.~Boyd and L.~Vandenberghe, \emph{Convex Optimization}.\hskip 1em plus 0.5em
  minus 0.4em\relax Cambridge University Press, 2004.

\bibitem{cvx}
M.~Grant and S.~Boyd, ``{CVX}: Matlab software for disciplined convex
  programming,'' \url{http://cvxr.com/cvx}, May 2010.

\bibitem{Palomar2003a}
D.~Palomar, J.~Cioffi, and M.~Lagunas, ``Joint {Tx}-{Rx} beamforming design for
  multicarrier {MIMO} channels: a unified framework for convex optimization,''
  \emph{{IEEE} Trans. Signal Process.}, vol.~51, no.~9, pp. 2381--2401, 2003.

\bibitem{Tuy1998a}
H.~Tuy, \emph{Convex analysis and global optimization}.\hskip 1em plus 0.5em
  minus 0.4em\relax Kluwer Academic Publishers, 1998.

\bibitem{Wiesel2008a}
A.~Wiesel, Y.~Eldar, and S.~Shamai, ``Zero-forcing precoding and generalized
  inverses,'' \emph{{IEEE} Trans. Signal Process.}, vol.~56, no.~9, pp.
  4409--4418, 2008.

\bibitem{Tarighat2005a}
A.~Tarighat, M.~Sadek, and A.~Sayed, ``A multi user beamforming scheme for
  downlink {MIMO} channels based on maximizing signal-to-leakage ratios,'' in
  \emph{Proc.~ICASSP'05}, 2005, pp. 1129--1132.

\bibitem{Peel2005a}
C.~Peel, B.~Hochwald, and A.~Swindlehurst, ``A vector-perturbation technique
  for near-capacity multiantenna multiuser communication---part {I}: Channel
  inversion and regularization,'' \emph{{IEEE} Trans. Commun.}, vol.~53, no.~1,
  pp. 195--202, 2005.

\bibitem{Fuchs2007a}
M.~Fuchs, G.~D. Galdo, and M.~Haardt, ``Low-complexity space-time-frequency
  scheduling for {MIMO} systems with {SDMA},'' \emph{{IEEE} Trans. Veh.
  Technol.}, vol.~56, no.~5, pp. 2775--2784, 2007.

\bibitem{Rhee2000a}
W.~Rhee and J.~Cioffi, ``Increase in capacity of multiuser {OFDM} system using
  dynamic subchannel allocation,'' in \emph{Proc.~IEEE VTC-Spring'00}, 2000,
  pp. 1085--1089.

\end{thebibliography}

\begin{IEEEbiography}[{\includegraphics[width=1in,height=1.25in,clip,keepaspectratio]{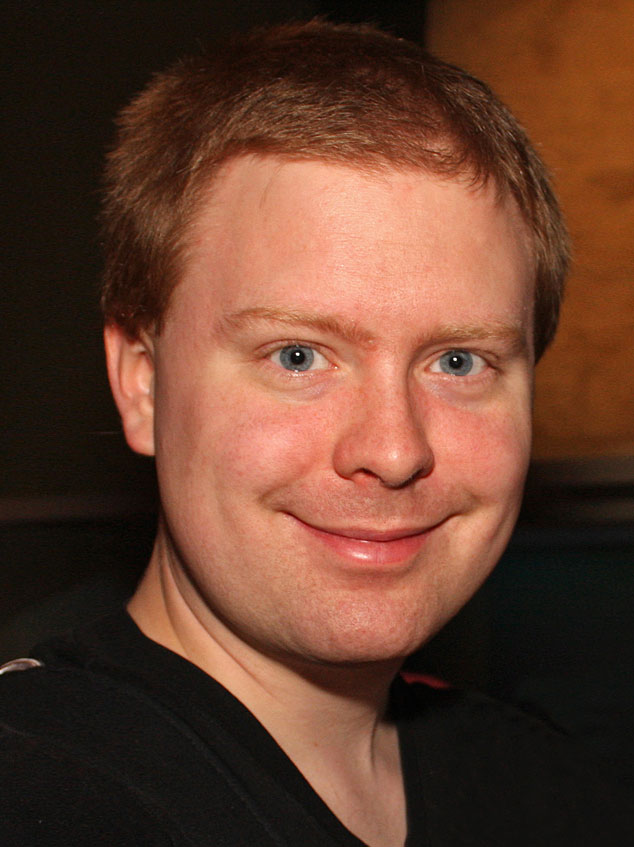}}]{Emil Bj\"ornson}
(S'07) was born in Malm\"o, Sweden,
in 1983. He received the M.S.~degree in engineering
mathematics from Lund University, Lund,
Sweden, in 2007. He is currently working toward the
Ph.D. degree in telecommunications at the Signal
Processing Laboratory, KTH Royal Institute of
Technology, Stockholm, Sweden.

His research interests include multiantenna communications,
multicell resource allocation, feedback
quantization, estimation theory, stochastic signal processing,
and mathematical optimization.

Dr.~Bj\"ornson received a Best Paper Award at the 2009 International Conference
on Wireless Communications and Signal Processing (WCSP 2009) for his
work on multicell MIMO communications.
\end{IEEEbiography}

\begin{IEEEbiography}[{\includegraphics[width=1in,height=1.25in,clip,keepaspectratio]{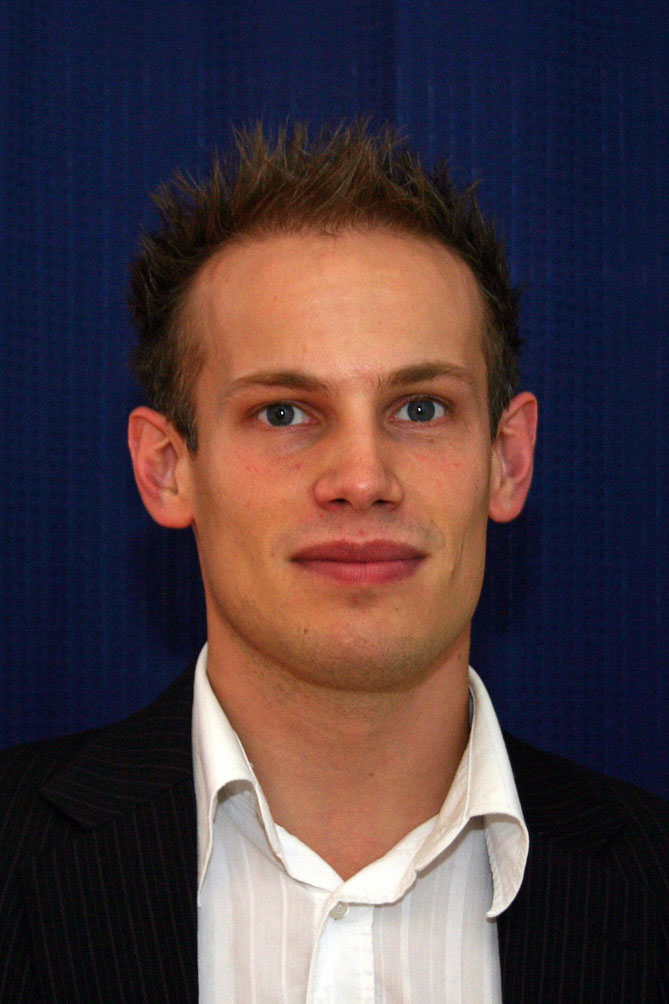}}]{Niklas Jald\'en}
(S'04-M'09) was born in G\"avle,
Sweden, in 1978. He received the M.S.~degree in
electrical engineering from the KTH Royal Institute
of Technology, Stockholm, Sweden, in 2004.
He received the Tech.~Lic.~and Ph.D.~degrees in
telecommunications from the same university, in
2007 and 2010, respectively.

He has since joined Ericsson Research, Stockholm,
Sweden. His research interests include
wireless channel modeling.

Dr.~Jald\'en received a Best Paper Award at the 2006
European Conference on Antennas and Propagation (EuCAP) for his work on
antenna patterns and channel measurement.
\end{IEEEbiography}

\begin{IEEEbiography}[{\includegraphics[width=1in,height=1.25in,clip,keepaspectratio]{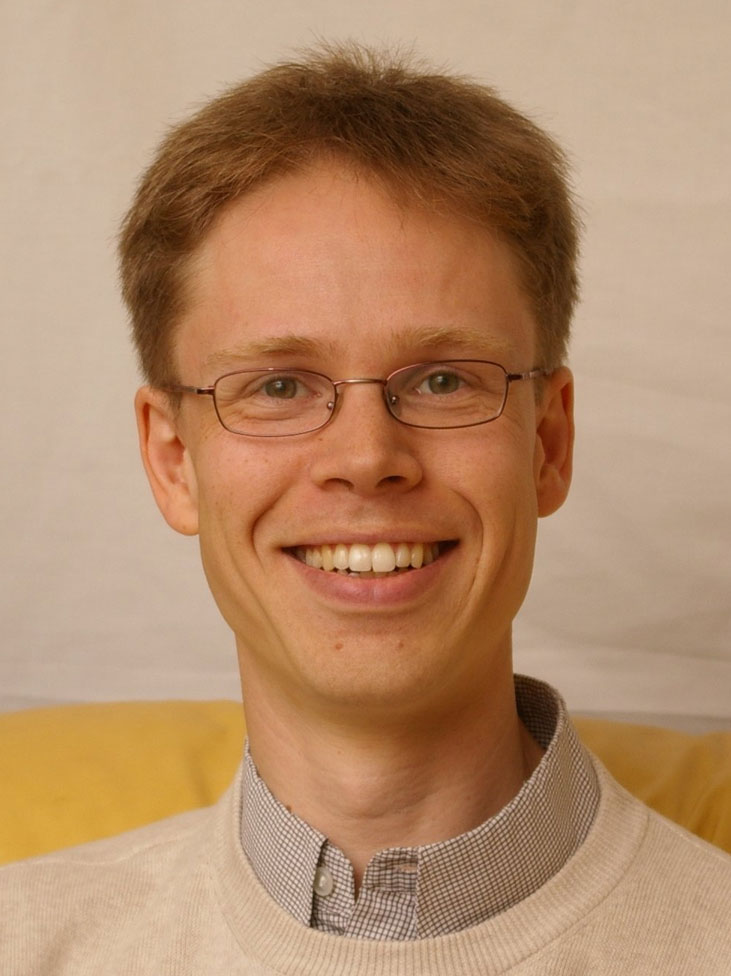}}]{Mats Bengtsson}
(M'00-SM'06) received the M.S.~degree in computer science from Link\"oping University,
Link\"oping, Sweden, in 1991, and the Tech.~Lic.~and Ph.D.~degrees in electrical engineering from
the KTH Royal Institute of Technology, Stockholm,
Sweden, in 1997 and 2000, respectively.

From 1991 to 1995, he was with Ericsson Telecom
AB Karlstad. He currently holds a position as Associate
Professor with the Signal Processing Laboratory,
School of Electrical Engineering, KTH. His
research interests include statistical signal processing
and its applications to communications, multiantenna processing, radio resource
management, and propagation channel modeling.

Dr.~Bengtsson served as an Associate Editor for the IEEE TRANSACTIONS ON
SIGNAL PROCESSING 2007-2009 and is a member of the IEEE SPCOM Technical
Committee.
\end{IEEEbiography}

\begin{IEEEbiography}[{\includegraphics[width=1in,height=1.25in,clip,keepaspectratio]{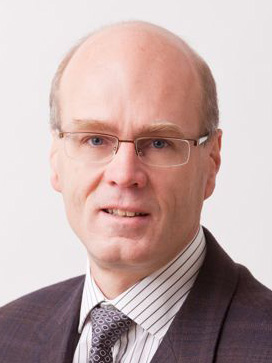}}]{Bj\"orn Ottersten}
(S'87-M'89-SM'99-F'04) was born in Stockholm, Sweden, in 1961. He
received the M.S. degree in electrical engineering and applied
physics from Link\"oping University, Link\"oping, Sweden, in 1986
and the Ph.D. degree in electrical engineering from Stanford
University, Stanford, CA, in 1989.

He has held research positions at the Department of Electrical
Engineering, Link\"oping University; the Information Systems
Laboratory, Stanford University; and the Katholieke Universiteit
Leuven, Leuven, Belgium. During 1996-1997, he was Director of
Research at ArrayComm Inc., San Jose, CA, a start-up company based
on Ottersten's patented technology. In 1991, he was appointed
Professor of Signal Processing at the KTH Royal Institute of
Technology, Stockholm, Sweden. From 2004 to 2008, he was Dean of the
School of Electrical Engineering at KTH, and from 1992 to 2004 he
was head of the Department for Signals, Sensors, and Systems at KTH.
He is also Director of security and trust at the University of
Luxembourg. His research interests include wireless communications,
stochastic signal processing, sensor array processing, and
time-series analysis.

Dr. Ottersten has coauthored papers that received an IEEE Signal
Processing Society Best Paper Award in 1993, 2001, and 2006. He has
served as Associate Editor for the IEEE TRANSACTIONS ON SIGNAL
PROCESSING and on the Editorial Board of the IEEE Signal Processing
Magazine. He is currently Editor-in-Chief of the EURASIP Signal
Processing Journal and a member of the Editorial Board of the
EURASIP Journal of Advances in Signal Processing. He is a Fellow of
IEEE and EURASIP. He is one of the first recipients of the European
Research Council advanced research grant.
\end{IEEEbiography}

\end{document}